# Direct Detection of Lyman Continuum Escape from Local Starburst Galaxies with the Cosmic Origins Spectrograph

Claus Leitherer

*Space Telescope Science Institute[1], 3700 San Martin Drive, Baltimore, MD 21218, USA*

*leitherer@stsci.edu*

Svea Hernandez

*Radboud University Nijmegen, Department of Astrophysics, P.O. Box 9010, M.S. 62,*

*6500 GL Nijmegen, The Netherlands*

*s.hernandez@astro.ru.nl*

Janice C. Lee

*Space Telescope Science Institute[1], 3700 San Martin Drive, Baltimore, MD 21218, USA*

*jlee@stsci.edu*

M. S. Oey

*University of Michigan, Astronomy Department, 311 West Hall, Ann Arbor, MI 48109,*

*USA*

*msoey@umich.edu*

---

**Abstract**

We report on the detection of Lyman continuum radiation in two nearby starburst galaxies. Tol 0440-381, Tol 1247-232 and Mrk 54 were observed with the Cosmic Origins Spectrograph onboard the Hubble Space Telescopes. The three galaxies have radial velocities of ~13,000 km s$^{-1}$, permitting a ~35 Å window on the restframe Lyman continuum shortward of the Milky Way Lyman edge at 912 Å. The chosen instrument configuration using the G140L grating covers the spectral range from 912 to 2,000 Å. We developed a dedicated background subtraction method to account for temporal and spatial background variations of the detector, which is crucial at the low flux levels around 912 Å. This modified pipeline allowed us to significantly improve the statistical and systematic detector noise and will be made available to the community. We detect Lyman continuum in all three galaxies. However, we conservatively interpret the emission in Tol 0440-381 as an upper limit due to possible contamination by geocoronal Lyman series lines. We determined the current star-formation properties from the far-ultraviolet continuum and spectral lines and used synthesis models to predict the Lyman continuum radiation emitted by the current population of hot stars. We discuss the various model uncertainties such as, among others, atmospheres and evolution models. Lyman continuum escape fractions were derived from a comparison between the observed and predicted Lyman continuum fluxes. Tol 1247-232, Mrk 54 and Tol 0440-381 have absolute escape fractions of (4.5 ± 1.2)%, (2.5 ± 0.72)% and <(7.1 ± 1.1)%, respectively.

# 1. Introduction

There is profound theoretical and observational interest in measuring the Lyman continuum escape fraction $f_{esc}$, i.e., the fraction of hydrogen-ionizing photons produced by young massive stars that evade absorption and travel various distances from the sites of gas and dust from which they originated, and in identifying the physical conditions that favor this "escape" (Benson et al. 2013). Studying $f_{esc}$ is critical for determining the sources responsible for ionization of gas on scales spanning from galactic disks to the Universe. We still do not know with certainty the properties of the sources that give rise to the diffuse, warm ionized medium within galaxies, the halo gas surrounding galaxies, and the reionization of the intergalactic medium (IGM) at redshift $z > 6$.

The warm ionized medium is a major component of the interstellar medium (ISM) in galaxies (Mathis 2000) and investigating its topology, composition, ionization conditions and relation to the other phases of the ISM is needed for a complete understanding of the evolution of the ISM in galaxies and its impact in the IGM. Massive OB stars in galaxies likely represent the main ionizing source (e.g., Haffner et al. 2009), although there is considerable debate regarding the origin of the ionizing radiation and the significance of density- versus ionization bounded H II regions (Oey et al. 2007). Pellegrini et al. (2012) found that on the order of 40% of ionizing photons leak out of density-bounded H II regions, sufficient to power the warm interstellar medium in the Magellanic Clouds.

Halo gas connects the baryon-rich IGM to the star-forming disks of galaxies (Putman et al. 2012). Radiation escaping from the disks ionizes and heats the halo gas. Hα observations of Galactic halo clouds suggest that a few percent of the Milky Way's ionizing photons escape normal to the disk (Bland-Hawthorn & Maloney 1999). For other disk galaxies, available $f_{\rm esc}$ values are mainly model-based, and the predictions are in the range of a few percent (e.g., Fernandez & Shull 2011).

Lyman continuum radiation from star formation appears to constitute an increasingly important ingredient of the ionizing background at large $z$ and may dominate the radiation field at the epoch of H I reionization (Duncan & Conselice 2015). Estimating the contribution from star formation empirically and comparing it to that of QSO's represents a fundamental goal for observational cosmology. If AGN were capable of achieving H I reionization at $z > 6$, their properties would have to be rather extreme in terms of the faint end of the luminosity function, which seems to be inconsistent with observations. Therefore AGN are not likely to be responsible for the reionization of H I (Fontanot et al. 2012).

Hot, massive stars are detected in galaxies at $z$ up to ~7 using proxies such as blue galaxy colors (Bouwens et al. 2012), warm dust emission (Barger et al. 2012), nebular emission lines (Stark et al. 2013), or directly from signatures like stellar absorption lines (Cassata et al. 2015). This suggests a plethora of ionizing photons in high-$z$ star-forming galaxies. Do these photons penetrate the galaxy ISM and escape into the IGM? This makes $f_{\rm esc}$ the key quantity in understanding the feedback from galaxies on the IGM at all redshifts, and it is a particularly important input parameter in reionization models and cosmological simulations of the early universe (Benson et al. 2013).

Bergvall et al. (2013) summarized the observational efforts to detect escaping Lyman continuum radiation in star-forming galaxies at all redshifts. Focusing on *direct* detections (as opposed to indirect inferences via, e.g., absorption-line strengths (Heckman et al. 2001), pre-HST attempts to detect the Lyman continuum in local star-forming galaxies were made with HUT. Leitherer et al. (1995) found relatively loose upper limits to the escape fractions in four galaxies of order 1 to 10%. Instrumental effects could push these limits even higher (Hurvitz et al. 1997). HUT was largely superseded by the more powerful Far-Ultraviolet Spectroscopic Explorer (FUSE; Moos et al. 2000), whose capabilities permitted the comprehensive study of Leitet et al. (2013). Lately, the Cosmic Origins Spectrograph (COS; Osterman et al. 2011; Green et al. 2012) aboard the Hubble Space Telescope (HST) is transforming the field by enabling Lyman continuum observations in local star-forming galaxies at high spectral resolution and signal-to-noise. Borthakur et al. (2014) reported the first *detection* of Lyman continuum emission in a local star-forming galaxy with COS. The derived relative and absolute escape fractions are 21% and 1%, respectively. Very recently, Izotov et al. (2016) found $f_{abs} = (7.8 \pm 1.1)\%$ in the local star-forming galaxy J0925+1403 with COS. This value is one of the highest determined at any redshift.

Attempts to detect escaping Lyman continuum radiation from star forming galaxies beyond the local universe fall into two categories. At $z \approx 1$, space based ultraviolet (UV) imaging or spectroscopy around the Lyman break has been used to infer escaping ionizing radiation (e.g., Bridge et al. 2010; Rutkowski et al. 2016). No conclusive detection has been made to date (Siana et al. 2015). At higher redshift ($z \approx 2 - 3$), both space-based and ground-based observations are feasible. Evidence of Lyman continuum emission must be taken with care due to possible contamination by low-redshift interlopers (Mostardi et al. 2015). Nevertheless, Lyman continuum radiation has been detected in about 10% of the galaxies surveyed (Siana et al. 2015), with escape fractions ranging between 5% and 30% for Lyman-break galaxies and Lyman-α emitters, respectively (Nestor et al. 2013). Vanzella et al. (2016) reported an exceptional case of a star-forming galaxy at $z = 3.2$ whose escape fraction was found to exceed 50%.

Measuring $f_{esc}$ at low versus high $z$ offers distinct advantages and challenges. The morphology of local galaxies can be studied at high spatial resolution, thereby providing unique insight into the structure of the ISM and the star-formation sites. Panchromatic ancillary data are typically abundant for nearby galaxies, which permit multi-pronged approaches to constrain the stellar and ISM properties. On the other hand, the combination of spectrograph apertures and the spatial extent of the galaxies limit the solid angle probed by the observations, and the observations may not be representative of the overall $f_{esc}$. Ironically, the advent of numerous $8 - 10$ m class telescopes combined with multi-object spectrographs has made it easier to collect significant numbers of restframe UV spectra of optically faint galaxies at $2 < z < 5$ (e.g., Nestor et al. 2013) than to obtain individual UV spectra of nearby galaxies from space. Furthermore, observations at high $z$ benefit from the existence of numerous deep surveys which have produced well defined galaxy samples whose properties permit careful elimination of biases and selection effects. In contrast, the local selection is often driven by UV photon starvation: even the most sensitive UV spectrographs can access only the brightest star-forming galaxies. One distinct advantage of local galaxies needs to be highlighted: absorption of escaping Lyman photons by the IGM along the line of sight is negligible and no modeling of the IGM radiative transfer is necessary for interpreting the data. Observations of the Lyman continuum at z > 3 experience significant attenuation by intergalactic neutral hydrogen over cosmological distances along the line of sight (Inoue et al. 2014), and observations of the galactic Lyman continuum must be interpreted in terms of both galactic and IGM properties.

Deharveng et al. (2001), Bergvall et al. (2006), Grimes et al. (2007), Grimes et al. (2009), and Leitet et al. (2013) analyzed FUSE spectra of local star-forming galaxies to search for escaping Lyman continuum radiation. The three galaxies discussed here (Tol 0440-381, Tol 1247-232, Mrk 54) were included in several of these studies. The conclusions reached by the cited papers are not all in agreement, with some studies reporting Lyman continuum detections and others deriving lower limits which are inconsistent with the detections for the same galaxies. The discrepant results are largely due to large systematics affecting the background subtraction of FUSE data at low source count rates (Leitet et al. 2011). Depending on the extraction and background model, detections may be missed or spurious data may be interpreted as detections. The most comprehensive analysis by Leitet et al. (2013) resulted in at least one firm detection of $f_{esc}$ = 2.4% in Tol 1247-232. The systematic uncertainties associated with faint sources observed with FUSE motivated us to employ COS for independent verification. COS has comparable sensitivity but has lower background noise. Equally important, its background noise is less temporally variable than that of FUSE, which is one major factor for the FUSE uncertainties. In this work we attempt a robust measurement of $f_{esc}$ in a carefully chosen sample of local starburst galaxies using COS. Our targets were pre-selected from the FUSE archive and test previous observations performed with FUSE.

This paper is organized as follows: In Section 2 we introduce and describe the target galaxies. The observations and the data reduction are discussed in Section 3. The galaxy spectra are presented in Section 4. In Section 5 we determine the star-formation properties. The Lyman continuum escape fractions are derived in Section 6, and the overall conclusions are presented in Section 7. An Appendix provides details on the data reduction pipeline.

## 2. Target galaxies

COS is capable of collecting useful spectra at wavelengths down to 912 Å (McCandliss et al. 2010). Its effective area of ~10 cm$^2$ is comparable to that of FUSE below 1000 Å. In addition, the COS detector background at these wavelengths is 2 – 3 times lower than the FUSE detector background. Therefore, COS exposure times for observations of detector background limited sources can be much shorter than in the case of FUSE – at the price of 10 times lower spectral resolution if the COS G140L grating is used.

In order to take advantage of the far-UV capabilities of COS, we modeled the trade-off between choosing brighter (and therefore closer) sources and the wavelength of the Lyman-break in the observed frame. The sweet spot occurs at $cz \approx 10{,}000$ to $15{,}000$ km s$^{-1}$. At lower velocity, the redshift window becomes unacceptably small since significant detections of the Lyman continuum require binning over some wavelength range. On the other hand, at higher redshift suitable candidates at sufficiently high flux levels become very scarce. We imposed a minimum flux level at 1000 Å of $F_\lambda(1000) > 1 \times 10^{-14}$ erg s$^{-1}$ cm$^{-2}$ Å$^{-1}$ for our candidate selection in order to keep the COS exposure times reasonable. We then queried the FUSE archive for suitable bright candidates in this redshift regime and identified the three brightest nuclear starburst galaxies. The central concentration of the UV light in nuclear starbursts minimizes aperture losses in COS with respect to FUSE whose entrance aperture is significantly larger. Table 1 gives some basic information for these galaxies: the most commonly used name (column 1), galaxy class
(column 2), Galactic foreground extinction (column 3), velocity *cz* (column 4), distance
*D* (column 5), absolute *B* magnitude $M_B$ (column 6), and the oxygen abundance
12+log(O/H) (column 7). All entries were taken from Leitet et al. (2013), except for *D*,
which we adopted from the NASA Extragalactic Database (NED). *D* is based on a
velocity-field model including the influence of the Virgo cluster, the Great Attractor, and
the Shapley supercluster (Mould et al. 2000). In addition to meeting the previous
selection criteria, the three galaxies have low Galactic foreground extinction (minimizing
Galactic $H_2$ contamination) and have no evidence of an AGN. The existing UV
spectroscopy from FUSE, HST and the International Ultraviolet Explorer (IUE) indicates
a hot-star dominated population capable of producing ample ionizing photons. The
question then is what fraction of these photons can escape from their galaxy host and
whether there are conditions that favor escape.

## 3. Observations and data reduction

The spectra of the three galaxies were obtained in program 13325
(http://www.stsci.edu/cgi-bin/get-proposal-info?id=13325&observatory=HST) during
several visits between December 2013 and May 2014. Table 2 gives the observation log.
The individual visits as designated by their archival identifiers are in column 1 of
Table 2. The galaxies were acquired through the primary science aperture (PSA) with a
standard NUV imaging acquisition using the target coordinates listed in columns 3 and 4
of the table. The onboard acquisition mechanism then centered the COS aperture on the
brightest region found within the $4 \times 4$ arcsec$^2$ field of view of the acquisition image.

After successful centering, spectra were obtained in TIME-TAG mode with the G140L grating of COS at a central wavelength setting of 1280 Å. This setting provides scientifically useful spectral coverage between 912 and 2200 Å, with a gap between 1165 and 1280 Å. The spectral region at wavelengths below 1165 Å is recorded on Segment B, and the region above 1280 Å on Segment A. We used the default number of four focal plane positions (FP-POS) to reduce fixed pattern noise. Therefore each individual exposure is broken into four sub-exposures with central wavelength settings offset by 20 Å. Columns 5 and 6 of Table 2 give the start times and durations of the exposures at each visit, respectively. The exposure times range between 18 and 28 ksec. Note that data sets LCAA10010, LCAA11010 and LCAA12010 were not used in this analysis. They contain low signal-to-noise data obtained in a visit suffering a guide-star acquisition failure. Instead we used LCAA51010 and LCAA52010, which contain the repeat observations of the failed visit. The nominal point-source resolving power of the G140L grating is $R \approx 2000$ at 1000 Å. However, this resolution is significantly reduced depending on the morphology of an extended source encompassed by the PSA. The circular PSA has 2.5″ diameter, which is substantially smaller than the $30 \times 30$ arcsec$^2$ FUSE aperture. This results in aperture losses for COS observations of extended galaxies compared with FUSE. Since the UV light of the program galaxies is highly concentrated, the COS fluxes are only lower than the FUSE fluxes by less than typically a factor of 2. More importantly, the smaller PSA leads to much reduced sky backgrounds, which is a crucial aspect for these particular observations. Observations in the satellite-UV are rarely sky-background limited in the continuum, and neither COS nor FUSE permit simultaneous object and sky measurements by default. The most significant sky background sources are geocoronal emission lines (Feldman et al. 2001), which can be identified and removed at longer wavelengths. Near the Lyman limit, however, geocoronal emission from the highest Lyman and other lines can contribute significantly to the total observed counts. This contribution is minimized by observations through a small aperture, such as the PSA.

We retrieved the individual data sets for each galaxy from the Mikulski Archive for Space Telescopes (MAST) and initially processed them on-the-fly with version 2.21 of the CalCOS pipeline. CalCOS is a series of modules which processes the data for detector noise, thermal drifts, geometric distortions, orbital Doppler shifts, count-rate non-linearity, and pixel-to-pixel variations in sensitivity. A standard wavelength scale is applied using the onboard wavelength calibration. The final products are one-dimensional, flux-calibrated, heliocentric-velocity corrected, and combined spectra.

The current background correction performed by CalCOS 2.21 uses a BACKCORR module that estimates the background contribution to the extracted spectrum and subtracts it from the science spectrum itself. These estimates are based on computations done using two predefined regions external to the spectral extraction region specified in the XTRACTAB. The actual background at the target location can differ from the background at the locations used for computing the background estimates which can lead to the over- or under-subtraction of the background/dark current to the final spectrum. An optimized and accurate background correction is needed when studying faint flux levels below 1150 Å. For this work we developed the algorithm described in Appendix A. The two major elements of our modified pipeline are a dedicated analysis of the pulse-height amplitude[2] and an improved background correction which is optimized for low source count rates. This algorithm is applicable to data obtained prior to February 2015 when the COS FUV spectra were moved to Lifetime Position 3 (LP3), with the exception of G130M/1055 and G130M/1096 which remained at Lifetime Position 2 (LP2). Due to the fact that the LP3 spectral location is sufficiently close to gain-sagged regions at Lifetime Position 1 (LP1), the COS instrument team has created a new algorithm to ensure good quality spectral extraction. This newly developed algorithm is used to calibrate data obtained at LP3 (COS STAN − Feb 2015). The steps and procedures described in Appendix A are only applicable to data taken at LP1 and/or LP2. The spectra discussed in the following were processed including the steps outlined in Appendix A.

The one-dimensional x1d files were individually analyzed for any quality issues. No issues were found. The x1d files were then further processed with an IDL pipeline derived from software originally written by the COS Guaranteed Time Observer (GTO) Team (Danforth et al. 2010). This software corrects the individual extracted spectra for major flat-field artifacts in the flux vector, such as ion repeller grid-wire shadows, and adjusts the errors and exposure times in the affected pixels accordingly. The decreased signal-to-noise at the detector edges is accounted for by de-weighting these regions. While some of these steps are also done in the default CalCOS pipeline, the IDL software employed leads to a significant improvement over the standard data reduction for low signal-to-noise data. The cross-correlated spectra were combined after interpolating the exposure-timed weighted flux vectors onto a common wavelength vector. The spectra were then resampled from their intrinsic dispersion of 0.0803 Å pix$^{-1}$ to 0.482 Å pix$^{-1}$, which corresponds to the nominal point-source resolution. The onboard wavelength calibration was checked by comparing the measured and laboratory wavelengths of the geocoronal O I lines at 1302.17/1304.86/1306.03 Å, when detected. The measured differences of less than one pixel indicate no significant offsets. Additional wavelength zero point shifts are caused by imperfect centering and the flux profiles of the galaxies in the PSA. We determined these shifts from the wavelength offsets of the Milky Way foreground absorption lines of C II λ1334.53, Si II λ1526.71, and Al II λ1670.79. We assumed these lines are unshifted and no Galactic high-velocity clouds are present (see Leitherer et al. 2011). Typical wavelength offsets of ~0.2 Å were found and removed. The Milky Way foreground lines are expected to be unresolved at the nominal COS G140L point-source resolution of R ≈ 2,000. We measured typical widths of ~2 Å in the three spectra, suggesting significant source structure and extension inside the PSA.

---

[2] The pulse-height amplitude characterizes the total charge in the electron cloud incident triggered by an incoming photon.

## 4. Galaxy spectra

We reproduce the spectra of Tol 0440-381, Tol 1247-232 and Mrk 54 which were processed through all previously discussed steps in Figure 1, 2 and 3, respectively. Identifications of spectral lines most commonly observed in star-forming galaxies are included in the figures. We refer the reader to Grimes et al. (2009) and Leitherer et al. (2011) for the atomic data and formation mechanisms of the listed lines. Note that not all labeled lines are actually detected in each galaxy. The line identifications at the top of each panel refer to redshifted lines intrinsic to the galaxies, whereas those at the bottom denote Galactic foreground absorptions at essentially zero redshift. As expected, the strongest Galactic absorption lines are observed in Tol 1247-232, which also has the highest foreground reddening. Geocoronal O I λ1300 emission is strong in Tol 0440-381 and Tol 1247-232. Geocoronal Lyman-series lines are of particular interest. While Ly-α is blocked by design, all other Lyman lines could in principle contaminate the spectra. Inspection of Figure 1 clearly indicates geocoronal Lyβ at 1026 Å in Tol 0440-381. We will address this issue in detail when discussing the uncertainties associated with the measurement of the Lyman continuum in the three galaxies.

The intrinsic spectral lines fall into three groups by virtue of their origin: interstellar, stellar photospheric, and stellar wind. Examples for these types are C II λ1335, S V λ1502, and C IV λ1550, respectively. A table summarizing the expected formation mechanism of most observed lines can be found in Leitherer et al. (2011; their Table 1). The most prominent stellar-wind lines are O VI λ1035 and C IV λ1550. These lines are important population diagnostics which will be used, together with other lines, to constrain the massive-star content. Si IV λ1400 has both a stellar and interstellar component, and its interpretation is more complex. N V λ1240 is widely used as a population indicator. Unfortunately, at the redshift of the galaxies this line coincides with geocoronal O I λ1300 and is not observable. Strong, broad He II λ1640 is detected in Tol 1247-232, which is consistent with its classification as a Wolf-Rayet (W-R) galaxy by Schaerer et al. (1999) based on the presence of broad He II λ4686 in the optical. C III λ1909 is seen in both Tol 0440-381 and Tol 1247-232 but not in Mrk 54, which is most likely a metallicity effect. The two former galaxies have Magellanic Cloud-like oxygen abundances, whereas the latter has near-solar oxygen abundance.

The spectra shortward and longward of the spectral gap display a pronounced dichotomy. At long wavelengths (the region recorded by Segment A), the spectrum has relatively few spectral features and closely follows a power law. In contrast, at short wavelengths (Segment B), strong spectral features cause heavy line-blanketing and the spectral energy distribution turns over and deviates from a power law. The Segment B data by themselves would make the measurement of the Lyman-break rather challenging since the location of the true continuum is elusive. Combining the Segment B and Segment A observations for a wider spectral coverage provides a much more reliable baseline for fitting the continuum. This was one of the motivations for choosing the COS G140L grating with its broader wavelength range over the more limited G130M grating.

Subsequent to the previous pipeline steps we corrected for Galactic foreground reddening with the $E(B-V)_{MW}$ values in Table 1 and the reddening law of Mathis (1990). Transformation from the observed to the restframe wavelengths was performed with the galaxy velocities in this table. Finally, we also generated rectified versions of the galaxy spectra, whose line-free regions were fitted with a multi-order polynomial and divided by the resulting continuum spectrum.

## 5. Star-formation properties

We can quantify the content of massive stars in the three galaxies by comparing the UV spectra to models calculated with the Starburst99 synthesis code (Leitherer et al. 1999; Vázquez & Leitherer 2005; Leitherer & Chen 2009; Leitherer et al. 2014). This technique has been used in the past and has become fairly standard (e.g., Leitherer et al. 2013). Therefore the description will be brief. We calculated a grid of synthetic spectra for multiple chemical compositions, stellar initial mass functions (IMF), star-formation histories, stellar atmospheres and stellar evolution models. The properties of our preferred model are: Kroupa-type IMF (Kroupa 2008, his eq. 1), which is identical to a Salpeter IMF at the high-mass end, continuous star formation over a period of 20 Myr, spherically extended, non-LTE, blanketed, expanding atmospheres for massive stars (Pauldrach et al. 1998; Hillier & Miller 1998, 1999; Gräfener et al. 2002; Hamann & Gräfener 2003, 2004), stellar evolution models with rotation (Ekström et al. 2012; Georgy et al. 2013) and solar chemical composition $Z_\odot$. Tol 0440-391 and Tol 1247-232 have chemical composition halfway between those available for the evolution models with rotation. Since the models with $Z_\odot$ have been tested much more extensively (there are no individual very metal-poor O stars in the Milky Way or the Magellanic Clouds), we opted for the solar models. We defer a comprehensive evaluation of the impact of the various ingredients and assumptions on the results to the following section, which discusses the escape of the Lyman continuum photons. In the following, we will refer to the adopted model as the "baseline model". The synthetic spectra of the baseline model were then fitted to the data *longward* of the Lyman break with the principal goal of determining the fraction of the Lyman continuum flux *shortward* of 912 Å escaping from the galaxies.

After correction for foreground reddening and redshift we compared the observed UV spectral energy distributions to the model predictions in order to determine the intrinsic dust attenuation. The spectral region between 1200 and 1800 Å is well approximated by a power law with an exponent β, and the difference between the observed and theoretical value of β permits an estimate of the intrinsic reddening $E(B-V)_{\text{intrinsic}}$ (Calzetti 2001). Together with $E(B-V)_{\text{MW}}$ (column 3 of Table 1) this then leads to the total reddening $E(B-V)_{\text{total}}$ for each galaxy. The values of $E(B-V)_{\text{total}}$ thus derived are in column 2 of Table 3. They are based on the extragalactic attenuation law of Calzetti et al. (2000), which provides the relation between the UV β and the optical reddening. This attenuation law has a shallower rise from the optical to the UV than, e.g., the Galactic law of Mathis (1990). It was derived empirically for a sample of local starburst galaxies with relatively low dust content. Its properties are thought to result from an extended, non-uniform distribution of massive, ionizing and less massive, non-ionizing stars. The former are responsible for powering the nebular lines and the latter for emitting the stellar UV continuum. The most massive stars are still surrounded by the natal dust clouds whereas winds and supernovae have evacuated the interstellar surroundings of the older, less massive stars. This provides a natural explanation of shape of the law and its different attenuation of stars and gas. In contrast to the stars, the gas is thought to be homogeneous, having a uniform opacity for the ionizing photons emitted by the massive stars.

      The Calzetti law does not extend to wavelengths shorter than 1200 Å. Therefore we supplemented her relation with the attenuation law found for a sample of star-forming galaxies observed with the Hopkins Ultraviolet Telescope (HUT) by Leitherer et al. (2002). This relation is very similar to the one of Calzetti and permits a smooth extension from 1200 Å down to near the Lyman break.

      After correction for total reddening, the spectra were compared to scaled synthetic spectra with luminosity as the scaling factor. As the UV luminosity

predominantly depends on the current star-formation rate (*SFR*) in star-forming galaxies, *SFR* can straightforwardly be derived for an assumed set of secondary parameters.

The values of *SFR* for the areas encompassed by the COS aperture in the three galaxies are in column 3 of Table 3. In column 4 of this table we list the corresponding number of hydrogen ionizing photons $N_{\mathrm{Lyman}}$. The predicted fluxes shortward of the Lyman edge, $F_{\lambda,\mathrm{pred}}(912^-)$, are given in column 5. The values are obtained by averaging the theoretical continuum over a 30 Å interval below 912 Å where the spectral energy distribution is essentially flat. The normalization was chosen such that the values in column 5 would be applicable to the observed spectra after correction for total reddening. Expressed equivalently, the theoretical monochromatic luminosities for a given *SFR* were divided by $4\pi D^2$.

In Figure 4, 5 and 6 we compare the observed and synthetic spectra obtained with the parameters summarized in Table 3 for Tol 0440-381, Tol 1247-232, and Mrk 54, respectively. The synthetic spectra are based on the library of Leitherer et al. (2010). They were degraded from their native resolution of ~0.4 Å to 2 Å in order to match the resolution of the data. The models match the continuum slopes between 1200 and 1800 Å. Care is required when inspecting shorter wavelengths, where strong blanketing by interstellar lines in the data makes it difficult to define the continuum at resolution of the reproduced figures. Likewise, at the longest wavelengths, low count rates and uncertainties of the flux calibration may affect the data quality. The major spectral features are generally reproduced as well if contamination by interstellar lines is accounted for. This applies to O VI $\lambda$1035, Si IV $\lambda$1400, and C IV $\lambda$1550. (Recall that N V λ1240 is heavily contaminated by geocoronal O I.) Overall, the spectra are typical of galaxies with strong starbursts, which Starburst99 models tend to reproduce rather well.

## 6. Lyman continuum escape fractions

The theoretical luminosity-calibrated spectra predict the intrinsic Lyman continuum flux prior to any attenuation by gas and dust. We use this prediction for a determination of the relative escape fraction $f_{rel}$ of Lyman continuum photons. $f_{rel}$ is obtained from a comparison of the dereddened observed fluxes; therefore it is a lower limit to the actual escaping Lyman radiation, which in addition is decreased by dust attenuation. We refer to the latter as the absolute escape fraction $f_{abs}$, which accounts for both gas and dust attenuation.

In Figure 7, 8 and 9 we provide a zoomed view of the spectral region around the Lyman edge in Tol 0440-381, Tol 1247-232 and Mrk 54, respectively. The data are the same as in the previous three figures, here with the error vectors included. The latter account for the statistical errors (i.e., the 1σ Poisson noise) as extracted from the x1d files. The statistical errors per resolution element *before binning* are below the flux levels longward of the Lyman break, but are higher shortward of the break. The data below 912 Å can be compared to the theoretically predicted continuum level, which we have marked by the thick horizontal line. The theoretical values are the result of averaging the models over an interval of 30 Å between 875 and 905 Å. The data plotted in the figures suggest detections for all three galaxies (but see below). In Table 4 we list the measured, averaged flux levels both for the reddening corrected, $F_{\lambda,dered}(912^-)$, and uncorrected, $F_{\lambda,obs}(912^-)$, fluxes in columns 2 and 3, respectively. $F_{\lambda,dered}(912^-)$ are the fluxes shown in Figure 7, 8 and 9. As we did for the theoretical spectra, we obtained the values of $F_{\lambda,\text{dered}}(912^-)$ and $F_{\lambda,\text{obs}}(912^-)$ by averaging the restframe Lyman continuum below 912 Å. The errors in the table were calculated from the statistical errors, taking into account the signal-to-noise gain by averaging over ~80 resolution elements. $f_{\text{rel}}$ can then straightforwardly be calculated from the ratio of $F_{\lambda,\text{dered}}(912^-)$ and $F_{\lambda,\text{pred}}(912^-)$. The resulting relative escape fractions are in column 4 of Table 4. Tol 1247-232 and Mrk 54 have $f_{\text{rel}}$ values of about 20%, whereas Tol 0440-381 has a higher value of 60%.

The absolute escape fractions are the more relevant quantity since they are a measure of, e.g., the photon supply for ionizing the IGM. However, they are dependent on the derived value of the dust attenuation, as well as the reddening law which was adopted. In our case, the adopted Calzetti law assumes a patchy foreground screen for the dust attenuation. This attenuation law does not relate the dust distribution to the distribution of the gas. The latter has the simplest opacity morphology, i.e., a uniform screen absorbing most ionizing photons. This idealized model does not consider alternatives in which the photon escape is the result of a certain covering factor in a picket fence model (Heckman et al. 2001). While either morphology (homogenous optically thin vs. picket fence optical thick) would lead to photon escape, their connection with the dust morphology could be rather different.

The absolute escape fractions $f_{\text{abs}}$ were obtained in the same manner as $f_{\text{rel}}$, but considering the ratio of $F_{\lambda,\text{obs}}(912^-)$ and $F_{\lambda,\text{pred}}(912^-)$, which accounts for the intrinsic dust obscuration. The results are in column 5 of Table 4. Leitherer et al. (2002) determined $A_{912}/E(B-V) \approx 15$ for the ratio of the total dust attenuation over the optical color excess in their attenuation law. The intrinsic dust reddening of the three galaxies, leads to attenuation factors making $f_{\text{abs}}$ a factor of $5 - 10$ smaller than $f_{\text{rel}}$. The values range from 2 to 7%, with Tol 0440-381 again having the largest escape fraction.

The errors quoted in Table 4 account for the statistical (Poisson) errors only. In addition, the data will be affected by errors introduced by both the sky and the detector background subtraction. We will address the detector background first. The detector background subtraction has been discussed in detail before. By creating a "superdark" we avoided introducing additional noise, and the method we employed is an improvement over the standard pipeline reduction. We note that Tol 0440-381 is the faintest of the three galaxies longward of the Lyman break and consequently has the largest detector background correction. Therefore any systematic error in the background would affect Tol 0440-381 the most.

The second source of systematic errors is the sky background. Far-UV observations are almost never dominated by the sky background (see Section 3), except possibly in discrete geocoronal lines. Therefore COS (like, e.g., FUSE) does not perform a simultaneous sky background in conjunction with the data collection. However, we can search for evidence of sky background contamination a posteriori. This issue is addressed by indicating the expected wavelengths of the higher geocoronal Lyman series lines in the wavelength restframe of each galaxy in Figure 7, 8 and 9. There is no detection of geocoronal Lyman-line emission in Tol 1247-232 and Mrk 54. In contrast, Lyman-$\beta$ and Lyman-$\gamma$ are clearly present in the spectrum of Tol 0440-381. This suggests that the merging higher Lyman series may contaminate the spectrum as well and that part, or all, of the rise of the data below 900 Å may be due to geocoronal Lyman lines. Analysis of archival FUSE spectra permits a quantitative assessment. Feldman et al. (2001) obtained

FUSE nightglow spectra covering the Lyman series from Lyman-β to Lyman-7. Their data suggest a combined strength of all Lyman lines from Lyman-7 up to the series limit of about 3% of Lyman-β. The geocoronal Lyman-β in the dereddened spectrum of Tol 0440-381 (see Figure 7) has a line flux of $2 \times 10^{-13}$ erg s$^{-1}$ cm$^{-2}$. If the Lyman decrement in this spectrum is the same as in the FUSE nightglow spectrum of Feldman et al., the expected flux of all Lyman lines from Lyman-7 upward is ~$6 \times 10^{-15}$ erg s$^{-1}$ cm$^{-2}$, with significant uncertainty. This is at least half the observed Lyman continuum level in Tol 0440-381. Since the other two galaxies show no detectable geocoronal Lyman-β, no significant contamination by geocoronal lines is expected in Tol 1247-232 and Mrk 54.

The derived escape fractions are > 3σ detections based on the statistical errors. Given the evidence of systematic errors affecting the Lyman continuum in Tol 0440-381, we conservatively consider the derived escape fractions in this galaxies as upper limits. Repeat observations, preferably at Dark Time, would be desirable to reject or confirm the observed Lyman continuum fluxes.

Next we address the model assumptions and uncertainties entering the determination of the escape fractions. Model atmospheres are the benchmark against which we compare the observed Lyman continuum. Here we use WM-Basic (Pauldrach et al. 1998) for O stars. WM-Basic solves both the radiative transfer and the hydrodynamics of stellar winds and predicts atmospheric fluxes accounting for non-LTE effects and line-blanketing. It is optimized for the UV where the major diagnostic metal lines formed in winds are located. Since there are no hydrogen lines of diagnostic value in this spectral region, only a simplified treatment of the hydrogen line modeling is implemented in WM-Basic. In particular, Stark broadening is not accounted for.

Therefore all Lyman lines formed in the photosphere (where collisional broadening matters) will be too weak in the models. Typically this affects all Lyman lines higher than Lyman-β. However, the computed atmospheric structure is not affected by this approximation, as demonstrated by Kaschinski et al. (2012). In Figure 10 we illustrate how the neglect of Stark broadening affects the Lyman lines in the models. The baseline population model shown has a resolution of 0.4 Å. The Lyman lines are clearly present at the expected wavelengths but are much narrower than would normally be observed in a population containing B stars. Evidently, this will not affect the determination of the continuum longward of ~930 Å. The only spectral region compromised by the inadequate treatment of the higher Lyman lines is between 912 and 920 Å, where the rise of the continuum in Figure 10 is an artifact due to the lack of Lyman blanketing. This region was not used for determining the Lyman break.

In order to study the impact of the major assumptions and modeling ingredients in a quantitative way, we vary age, IMF, chemical composition, model atmospheres and evolution models relative to the baseline model used for determining the escape fractions. This is addressed in Figures 11 – 15. In each of the figures we have plotted the spectral region around the Lyman break as predicted by the baseline model as well as the predictions resulting from variations of each of the mentioned parameters. The spectra are binned to a resolution of 10 Å. Figure 11 shows the spectra for different star-formation durations. Once star-formation has been ongoing for at least 10 Myr, an equilibrium between stellar birth and death of massive stars is reached, and the spectrum becomes age independent. Recall that star formation in our models is continuous, not instantaneous. Therefore the spectra are not very sensitive to the adopted age. Figure 12 illustrates the effect of the adopted IMF. Obviously a flat (rich in massive stars) IMF results in higher UV luminosities. This is simply a consequence of the normalization to the same total mass for all three spectra. Note that the Lyman break itself is much less affected: it decreases by ~0.2 dex from $\alpha = 2.3$ to 1.6. We adopted an IMF with $\alpha = 2.3$, which is widely observed with little evidence for variations (Bastian et al. 2010). Changes in chemical composition are addressed in Figure 13. We compare two models, one having solar chemical composition (the baseline model) and the other with $1/7^{th}$ solar. Contrary to expectation, the Lyman breaks for the two models are rather similar. This counterintuitive result is the consequence of two opposite trends: higher metallicity would favor a stronger Lyman break due to stronger line blanketing. However, stellar evolution at higher metallicity produces more hot W-R stars whose strong winds decrease the Lyman break. We emphasize that this is a property which is particularly pronounced in the adopted set of stellar evolution models.

In Figure 14 we return to the impact of the model atmospheres. Stars contributing to the spectral region around the Lyman break have temperatures hotter than ~20,000 K, corresponding to O, B and W-R stars. The baseline model adopts spherically extended, non-LTE, blanketed, expanding atmospheres for these stars (see Leitherer et al. 2014 for details). In particular, WM-Basic is used for O and early-B stars, and CMFGEN for W-R stars. Substituting classical, stationary LTE atmospheres for WM-Basic ("BaSel", Lejeune et al. 1998) leads to a stronger Lyman break. If, in addition, classical, stationary LTE atmospheres are used for W-R stars as well (dotted line in Figure 14), the Lyman increases even further since winds and non-LTE effects tend to decrease the Lyman break. The net effects are differences by up to a factor of 2. However, we do not consider these as model uncertainties since we know that hot stars have winds and stationary LTE atmospheres will underestimate the Lyman continuum.

Our final comparison concerns the impact of the stellar evolutionary tracks (Figure 15). The tracks adopted here include stellar rotation at a rate of 40% the break-up speed on the zero-age main-sequence (labeled v = 0.4 in Figure 15). The resulting spectrum can be compared to one computed with the same generation of tracks but with no rotation (v = 0) and to one with the previous generation of Geneva tracks released 1992 – 1994 ("1994" in Figure 15). Levesque et al. (2012), Levesque & Leitherer (2013) and Leitherer et al. (2014) provide extensive comparisons of the differences incurred in population synthesis models when switching between these evolutionary tracks. Models with rotation produce hotter O stars and larger number of W-R stars (at solar chemical composition), which explains the smaller Lyman break associated with the rotating models.

Among the modeling impacts discussed here, stellar evolution currently poses the largest uncertainty with possible future Lyman break revisions of up to a factor of 2. We also highlight the possibility of populations of massive binaries contributing to the UV flux (Eldridge 2012). Close massive binaries have different evolutionary paths which can produce ionizing radiation at amounts similar to single stars, but at later epochs (Stanway et al. 2016).

## 7. Discussion and conclusion

We detected escaping Lyman continuum radiation in Tol 1247-232 and Mrk 54 and derived a significant upper limit to the Lyman continuum in Tol 0440-381. The corresponding absolute escape fractions $f_{abs}$ are $(4.5 \pm 1.2)\%$, $(2.5 \pm 0.72)\%$ and $<(7.1 \pm 1.1)\%$ for Tol 1247-232, Mrk 54 and Tol 0440-381, respectively. For comparison, the relative escape fractions $f_{rel}$ are $(21.6 \pm 5.9)\%$, $(20.8 \pm 6.1)\%$ and $<(59.8 \pm 13)\%$.

We can compare our results to those determined by Leitet et al. (2013) who analyzed all useful data collected with FUSE for star-forming galaxies. The three target galaxies discussed here were part of their study. Leitet et al. found escape fractions of $f_{abs} = (2.4 \pm ^{0.9}_{0.8})\%$, $(2.5 \pm ^{1.2}_{1.0})\%$ and $< 0.3\%$ for Tol 1247-232, Mrk 54 and Tol 0440-381, respectively. Before comparing our and their results, differences in instrumentation and modeling approach must be taken into account. The FUSE entrance aperture has dimensions 30″ by 30″, which is more than 100 times larger than the circular 2.5″ COS entrance aperture. While we selected compact, nuclear starbursts, there may be some star formation occurring outside the COS aperture, which could contribute to the flux in the FUSE aperture. Additionally, Leitet et al. used IUE spectra to measure the UV spectral slope and to determine the reddening and the UV luminosity. Given these differences, we consider our and their result consistent within the statistical and systematic errors.

Various galaxy properties have been suggested as proxies for high Lyman continuum escape probabilities: low optical depth in low-ionization interstellar lines (Heckman et al. 2001), a dominant central object (Heckman et al. 2011), extreme [O III]/[O II] ratio (Jaskot & Oey 2013) or large Lyman-α equivalent width (Verhamme et al. 2015). The three target galaxies in this study fall into one or more of these categories, but not into all simultaneously. For instance, Tol 1247-232 and Mrk 54 are Lyman continuum leakers; while Tol 1247-232 shows strong Lyman-α emission, Mrk 54 has a damped Lyman-α absorption profile (Leitet et al. 2013). Likewise, Leitet et al.

found no clear correlation between the upper limits inferred from the absorption line optical depths and the escape fractions measured directly. Non-isotropic leakage of radiation may complicate the interpretation of these Lyman escape tracer.

Star-forming galaxies are thought to be the sources of reionization ending the "Dark Age" in the early universe (e.g., Haardt & Madau 2012). In order to fully ionize the universe by $z \approx 6$, escape fractions of ionizing radiation from galaxies of order 20% are needed (Robertson et al. 2013). Numerical simulations of galaxy formation predict a wide range of escape fractions but tend to be less than 20% (e.g., Paardekooper et al. 2015). This is largely the result of massive stars still being embedded in their birth sites when their production rate of ionizing photons peaks. Once these stars have cleared their surroundings, their ionizing luminosity falls precipitously and becomes insufficient for ionizing the IGM. Ma et al. (2016) studied the effect of binary evolution on the properties of the ionizing population. Mass transfer in binaries delays the production of ionizing radiation compared to single-star evolution, while leaving the overall luminosity largely unchanged. Therefore, more ionizing photons are available when stellar winds and supernovae have created holes and escape channels for the radiation. This boosts the escape fractions by factors of 3 – 6.

Observationally, the available data are still inconclusive. The IGM opacity precludes direct measurements of the escape fraction for $z > 3$. Around z = 2 – 3, observations are incomplete and plagued by selection effects. Local galaxies such as those presented in this paper are important training sets. Their proximity allows us to study them in detail and identify properties affecting Lyman continuum escape. The absolute escape fractions derived here are less than would be required if these galaxies were at cosmological redshift and had properties similar to those thought to be responsible for reionization. However, all three galaxies are dusty and suffer from significant dust attenuation in the UV. Gnedin et al. (2008) performed N-body + hydrodynamics simulations of the escape fraction of ionizing radiation in high-redshift galaxies. Their simulations accounted for the presence of different amounts of dust. Observationally, the amount of dust in galaxies at or near the epoch of reionization is unknown. Gnedin et al. found that realistic dust column densities would result in a difference of a factor of ~10 between the relative and absolute escape fractions. The relative escape fractions become relevant if the galaxies had little or no dust. Interestingly, the relative escape fractions in two galaxies are ~20%, resembling the value suggested as required for the reionization of the universe (Hartley & Ricotti 2016).

*Acknowledgments.* We thank Paul Feldman (JHU) for providing measurements of the geocoronal Lyman lines in a FUSE nightglow spectrum. An anonymous referee provided suggestions which helped improve the paper. Support for this work has been provided by NASA through grant number GO-13325 from the Space Telescope Science Institute, which is operated by AURA, Inc., under NASA contract NAS5-26555. This research has made extensive use of the NASA/IPAC Extragalactic Database (NED) which is operated by the Jet Propulsion Laboratory, California Institute of Technology, under contract with the National Aeronautics and Space Administration.

# Appendix A: Modifications to the CalCOS pipeline

**1. Background correction in CalCOS 2.21d using DARKFILE**

In order to optimize the background subtraction for low-count data, we modified CalCOS 2.21 to perform a dark subtraction using a two-dimensional superdark reference file. When the DARKFILE keyword is present in the primary header, and its value is other than N/A, CalCOS 2.21d subtracts the superdark image from the science exposure right before the 1-D extraction. The new dark-subtracted FLT is saved to a file with extension *darkcorrflt_a/b.fits* for future reference. The ERROR and DQ image extensions are copied directly from the *flt_a/b.fits* to the *darkcorrflt_a/b.fits* image. When using the DARKFILE method correction, we turned off BACKCORR (BACKCORR = OMIT) so that only the local dark current is subtracted from the spectral extraction region. Appendix Figure 1 illustrates the effects of the dark subtraction with the DARKFILE keyword set. After subtraction of the superdark file, the DARK-corrected FLT file has zero background level.

**2. Creation of the superdark (DARKFILE)**

In order to use the new background correction feature in CalCOS 2.21d, a darkfile must be provided. This reference file must be an image with the same dimensions as those of the science exposure. The keyword DARKFILE and the corresponding reference file name need to be present in the primary header of the science files to be calibrated. We followed the steps below for the creation of the darkfile.

During each cycle the STScI COS team executes an FUV Detector Dark Monitor program designed to obtain at least five dark exposures per week (EXPTIME= 1320s) to monitor the detector dark rate. In order to support Lifetime Position 3 and multiple High Voltage (HV) operations, starting in Cycle 22 the instrument team took an additional five exposures per week at the HV used for the G140L setting since this mode operates at a different HV than the other modes. We collected the dark exposures taken as part of the monitoring program and used them for the creation of the superdark. The COS FUV dark rate varies as a function of time. Appendix Figure 2 shows the measured COS FUV dark between 2010 and 2015, which includes the epoch when our observations were obtained. The data in the figure emphasize the need for contemporaneous dark measurements. Consequently, we considered only darks which were taken within the same month the science data were obtained. Furthermore, only dark exposures taken at the same HV as the one used in the science observations were retrieved. Our final superdarks were created using a total of typically 20 dark exposures.

**3. PHA Filtering**

The standard CalCOS 2.21 pipeline performs a Pulse Height Amplitude (PHA) correction on the COS FUV data using the PHACORR module. Since Pulse Height information is available for individual events when using FUV TIME-TAG mode, the PHACORR module compares the events' PHA to predefined thresholds stored in the PHATAB reference file. If the PHA values of such events are outside of these lower/upper level thresholds, the events are flagged with a DQ bit of 512 and omitted from the final spectrum. The predefined PHA thresholds in the PHATAB reference file have been chosen based on the properties of the detector itself; these properties change with continuing exposure to photons, an effect known as gain-sag. Additional PHA screening can contribute to the reduction of dark current counts in the COS FUV images (Ely et al.

2014). The current PHA thresholds used in the PHATAB reference file are: Lower Level Threshold (LLT) = 2, Upper Level Threshold (ULT) = 23. It has been shown before that decreasing the lower level threshold results in an increase of the background of ~7% (Sahnow et al. 2011). Additionally, it is known that background events, both internal and external to the detector, tend to have the lowest and highest pulse heights (see COS IHB Section 4.1.7). Specifying rejection thresholds (LLT and ULT) relative to the modal gain found in each science exposure (considering no gain-sag is present at the time of the observations) will restore the 7% increase. For present analysis, we selected new thresholds of LLT = 4 and ULT = 18. In Appendix Figure 3 we reproduce the Pulse Height Distribution (PHD) used in selecting the new PHA levels. These new PHA thresholds were applied to both the science and the dark exposures for consistency.

Appendix Figure 4 compares the mean dark rates estimated for the two dark exposure sets, one using the standard PHA thresholds (LLT = 2 and ULT = 23) and the other using the new PHA thresholds (LLT = 4 and ULT = 18). The figure demonstrates a clear ~25% decrease in the dark rate when using the new PHA thresholds. The same thresholds were used when calibrating both the dark and the science exposures.

## 4. Final preparation of the superdarks

After obtaining FLT files with the updated PHA thresholds (with units of counts/sec) for the individual dark exposures, we summed and divided them by the total number. The resulting superdark was then scaled to the science exposure. The scaling factor was calculated using the science frame and then multiplied by the superdark. Specifically, the scaling factor was estimated by obtaining an average dark rate from the background regions (defined in the XTRACTAB) in the science frames and dividing by the average dark rate (estimated using the same background regions as in the science) in the superdark itself: *scaling_factor = sci_dark_rate/super_dark_rate*.

When creating the superdark files, we faced the challenge of not having enough dark exposures to be able to characterize the dark structure in the detector. As mentioned earlier, the COS Dark Monitor only collects ~20 exposures per month. Because of this limitation, the superdark requires smoothing. This causes the reference files to lose some of their small-scale features, but their overall structure is still preserved. For our analysis, smoothing was done using a boxcar of size (10,100). Appendix Figure 5 shows an example of the cross-dispersion profiles of the science FLTs and the smoothed superdark.

**5. Resulting science data**

The use of the new version of CalCOS, 2.21d, along with the new PHATAB, modified the flux in the science exposures by ~100% or more around wavelengths between 920 and 970 Å. As an example, we show in Appendix Figure 6 and 7 the COS FUV Segment B smoothed calibrated spectrum for three different setups of the pipeline processing: using CalCOS 2.21 and standard PHATAB thresholds, CalCOS 2.21 and new PHATAB thresholds, and finally using the modified version of CalCOS (2.21d) with the new PHATAB values.

The COS throughput drops drastically at wavelengths below 1150 Å. This effect can be appreciated in Appendix Figure 6. Consequently, a precise background subtraction becomes more and more important. As expected, the difference between the three spectra in this figure increases with decreasing wavelengths, in particular below 1070 Å, where the instrument's throughput is the lowest. Small errors in the dark correction in regions where the instrument's throughput is noticeably lower than in the rest of the detector propagate as relatively large changes in the flux.

# REFERENCES

Barger, A. J., Wang, W.-H., Cowie, L. L., et al. 2012, ApJ, 761, 89

Bastian, N., Covey, K. R., & Meyer, M. R. 2010, ARA&A, 48, 339

Benson, A., Venkatesan, A., & Shull, J. M. 2013, ApJ, 770, 76

Bergvall, N., Leitet, E., Zackrisson, E., & Marquart, T. 2013, A&A, 554, 38

Bergvall, N., Zackrisson, E., Andersson, B.-G., et al. 2006, A&A, 448, 513

Bland-Hawthorn, J., & Maloney, P. R. 1999, ApJ, 510, L33

Borthakur, S., Heckman, T. M., Leitherer, C., & Overzier, R. A. 2014, Sci, 346, 216

Bouwens, R. J., Illingworth, G. D., Oesch, P. A., et al. 2012, ApJ, 754, 83

Bridge, C. R., Teplitz, H. I., Siana, B., et al. 2010, ApJ, 720, 465

Calzetti, D. 2001, PASP, 113, 1449

Calzetti, D., Armus, L., Bohlin, R. C., et al. 2000, ApJ, 533, 682

Danforth, C. W., Keeney, B. A., Stocke, J. T., Shull, J. M., & Yao, Y. 2010, ApJ, 720, 976

Deharveng, J.-M., Buat, V., le Brun, V., et al. 2001, A&A, 375, 805

Duncan, K., & Conselice, C. J. 2015, MNRAS, 451, 2030

Ekström, S., Eggenberger, P., Meynet, G., et al. 2012, A&A, 537, 146

Eldridge, J. J. 2012, MNRAS, 422, 794

Ely, J., et al. 2014, American Astronomical Society, AAS Meeting #224, #122.07

Feldman, P. D., Sahnow, D. J., Kruk, J. W., Murphy, E. M., & Moos, H. W. 2001, JGR, 106, 8119

Fernandez, E. R., & Shull, J. M. 2011, ApJ, 731, 20

Fontanot F., Cristiani S., & Vanzella E., 2012, MNRAS, 425, 1413

Georgy, C., Ekstrom, S., Eggenberger, P., et al. 2013, A&A, 558, 103

Gnedin, N. Y., Kravtsov, A. V., & Chen, H.-W. 2008, ApJ, 672, 765

Gräfener, G., Koesterke, L., & Hamann, W.-R. 2002, A&A, 387, 244

Green, J. C., Froning, C. S., Osterman, S., et al. 2012, ApJ, 744, 60

Grimes, J. P., Heckman, T., Aloisi, A., et al. 2009, ApJS, 181, 272

Grimes, J. P., Heckman, T., Strickland, D., et al. 2007, ApJ, 668, 891

Haardt, F., & Madau, P. 2012, ApJ, 746, 125

Haffner, L. M., Dettmar, R.-J., Beckman, J. E., et al. 2009, RvMP, 81, 969

Hamann, W.-R., & Gräfener, G. 2003, A&A, 410, 993

Hamann, W.-R., & Gräfener, G. 2004, A&A, 427, 697

Hartley, B., & Ricotti, M. 2016, MNRAS, submitted (arXiv:1602.06302)

Heckman, T. M., Borthakur, S., Overzier, R., et al. 2011, ApJ, 730, 5

Heckman, T. M., Sembach, K. R., Meurer, G. R., et al. 2001, ApJ, 558, 56

Hillier, D. J., & Miller, D. L. 1998, ApJ, 496, 407

Hillier, D. J., & Miller, D. L. 1999, ApJ, 519, 354

Hurwitz, M., Jelinsky, P., & Dixon, W. V. D. 1997, ApJ, 481, L31

Inoue, A. K., Shimizu, I., Iwata, I., & Tanaka, M. 2014, MNRAS, 442, 1805

Izotov, Y. I., Orlitová, I., Schaerer, D., et al. 2016, Nature, 529, 178

Jaskot, A. E., & Oey, M. S. 2013, ApJ, 766, 91

Kaschinski, C. B., Pauldrach, A. W. A., & Hoffmann, T. L. 2012, A&A, 542, 45

Kroupa, P. 2008, in Pathways Through an Eclectic Universe, ed. J. H. Knapen, T. J. Mahoney, & A. Vazdekis (San Francisco: ASP), 3

Leitet, E., Bergvall, N., Hayes, M., Linné, S., & Zackrisson, E. 2013, A&A, 553, A106

Leitet, E., Bergvall, N., Piskunov, N., & Andersson, B.-G. 2011, A&A, 532, A107

Leitherer, C., Chandar, R., Tremonti, C. A., Wofford, A., & Schaerer, D. 2013, ApJ, 772, 120

Leitherer, C., & Chen, J. 2009, New Astronomy, 14, 356

Leitherer, C., Ekström, S., Meynet, G., et al. 2014, ApJS, 212, 14

Leitherer, C., Ferguson, H. C., Heckman, T. M., & Lowenthal, J. D. 1995, ApJ, 454, L19

Leitherer, C., Li, I.-H., Calzetti, D., & Heckman, T. M. 2002, ApJS, 140, 303

Leitherer, C., Ortiz Otalvaro, P. A., Bresolin, F., et al. 2010, ApJS, 189, 309

Leitherer, C., Schaerer, D., Goldader, J. D., et al. 1999, ApJS, 123, 3

Leitherer, C., Tremonti, C. A., Heckman, T. M., & Calzetti, D. 2011, AJ, 141, 37

Lejeune, T., Cuisinier, F., & Buser, R. 1998, A&AS, 130, 65

Levesque, E. M., & Leitherer, C. 2013, ApJ, 779, 170

Levesque, E. M., Leitherer, C., Ekstrom, S., Meynet, G, & Schaerer, D. 2012, ApJ, 751, 61

Ma, X., Hopkins, P. F., Kasen, D., et al. 2016, MNRAS (ArXiv e-prints, arXiv:1601.07559)

Mathis, J. S. 1990, ARA&A, 28, 37

Mathis, J. S. 2000, ApJ, 544, 347

McCandliss, S. R., France, K., Osterman, S., et al. 2010, ApJ, 709, L183

Moos, H. W., Cash, W. C., Cowie, L. L., et al. 2000, ApJ, 538, L1

Mould, J. R., et al. 2000, ApJ, 529, 786

Mostardi, R. E., Shapley, A. E., Steidel, C. C., et al. 2015, ApJ, 810, 107

Nestor, D. B., Shapley, A. E., Kornei, K. A., Steidel, C. C., & Siana, B. 2013, ApJ, 765, 47

Nestor, D. B., Shapley, A. E., Steidel, C. C., & Siana, B. 2011, ApJ, 736, 18

Oey, M. S., Meurer, G. R., Yelda, S., et al. 2007, ApJ, 661, 801

Osterman, S., Green, J., Froning, C., et al. 2011, Ap&SS, 335, 257

Paardekooper, J.-P., Khochfar, S., & Dalla Vecchia, C. 2015, MNRAS, 451, 2544

Pauldrach, A. W. A., Lennon, M., Hoffmann, T. L., et al. 1998, in ASP Conf. Ser. 131, Properties of Hot Luminous Stars, ed. I. Howarth (San Francisco, CA: ASP), 258

Pellegrini, E. W., Oey, M. S., Winkler, P. F., et al. 2012, ApJ, 755, 40

Putman, M. E., Peek, J. E. G., & Joung, M. R. 2012, ARA&A, 50, 491

Robertson, B. E., Furlanetto, S. R., Schneider, E., et al. 2013, ApJ, 768, 71

Rutkowski, M. J., Scarlata, C., Haardt, F., et al. 2016, ApJ, 819, 81

Sahnow, D. J., Oliveira, C., Aloisi, A., et al. 2011, Proc. SPIE, 8145, 81450Q

Schaerer, D., Contini, T., & Pindao, M. 1999, A&AS, 136, 35

Siana, B., Shapley, A. E., Kulas, K. R., et al. 2015, ApJ, 804, 17

Stanway, E. R., Eldridge, J. J., & Becker, G. D. 2016, MNRAS, 456, 485

Stark, D. P., Schenker, M. A., Ellis, R., et al. 2013, ApJ, 763, 129

Vanzella, E, de Barros, S., Vasei, K., et al. 2016, ApJ, submitted (arXiv:1602.00688)

Vázquez, G. A., & Leitherer, C. 2005, ApJ, 621, 695

Verhamme, A., Orlitová, I., Schaerer, D., & Hayes, M. 2015, A&A, 578, A7

**Figures**

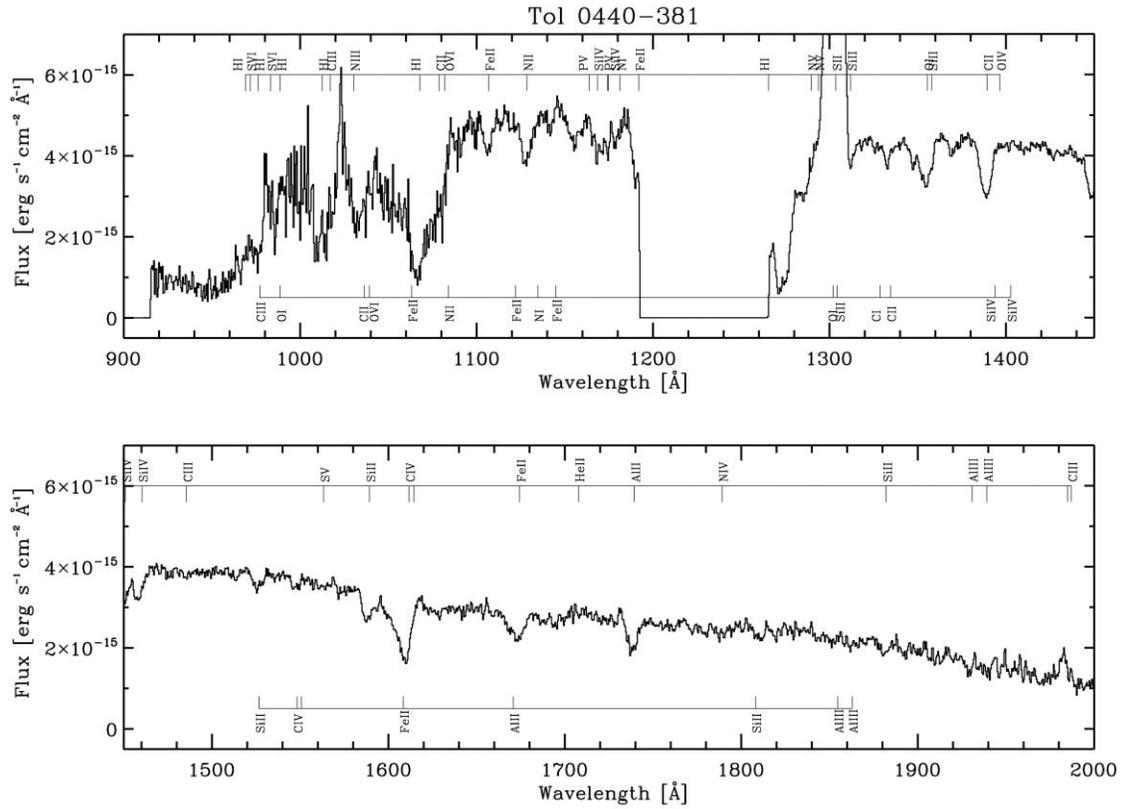

**Figure 1.** Observed UV spectrum of Tol 0440-381 from 900 to 2000 Å. The wavelength scale is in the observed frame; the fluxes are not corrected for reddening. The region between ~1190 and 1270 Å falls onto the detector gap. Identifications of intrinsic and Galactic foreground spectral features are given at the top and the bottom of each panel, respectively.

**Figure 2.** Same as Figure 1 but for Tol 1247-232.

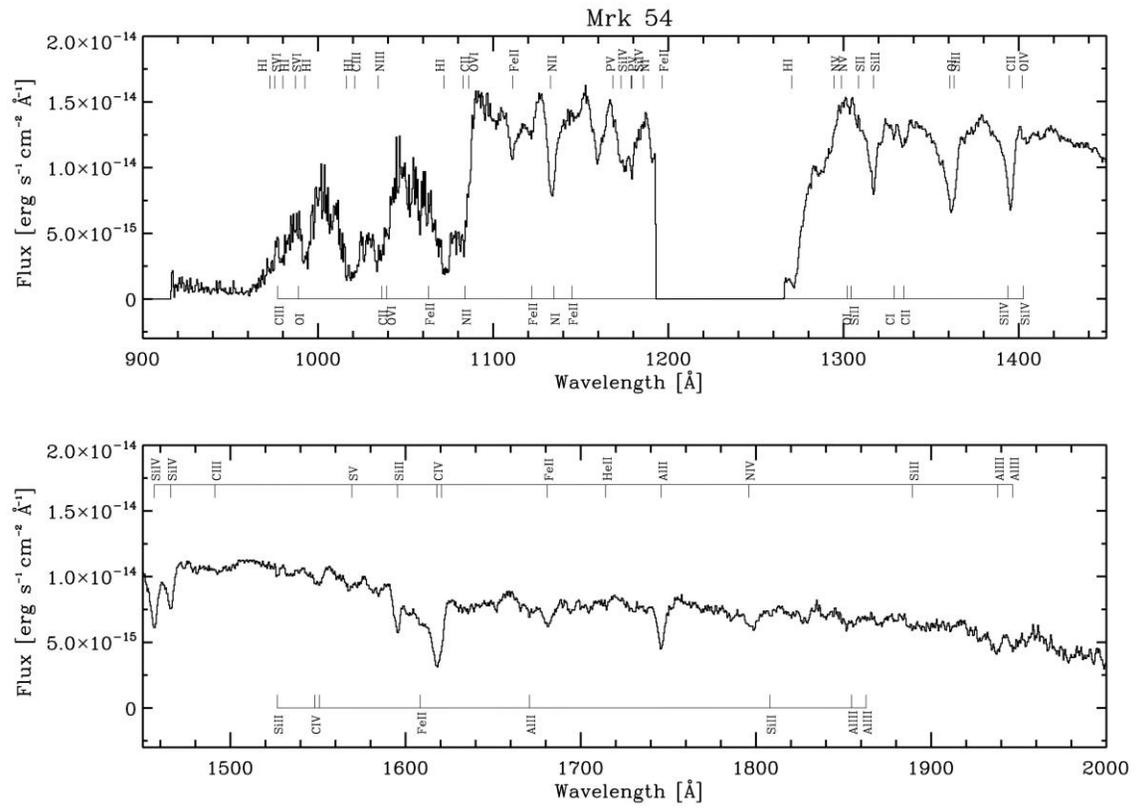

**Figure 3.** Same as Figure 1 but for Mrk 54.

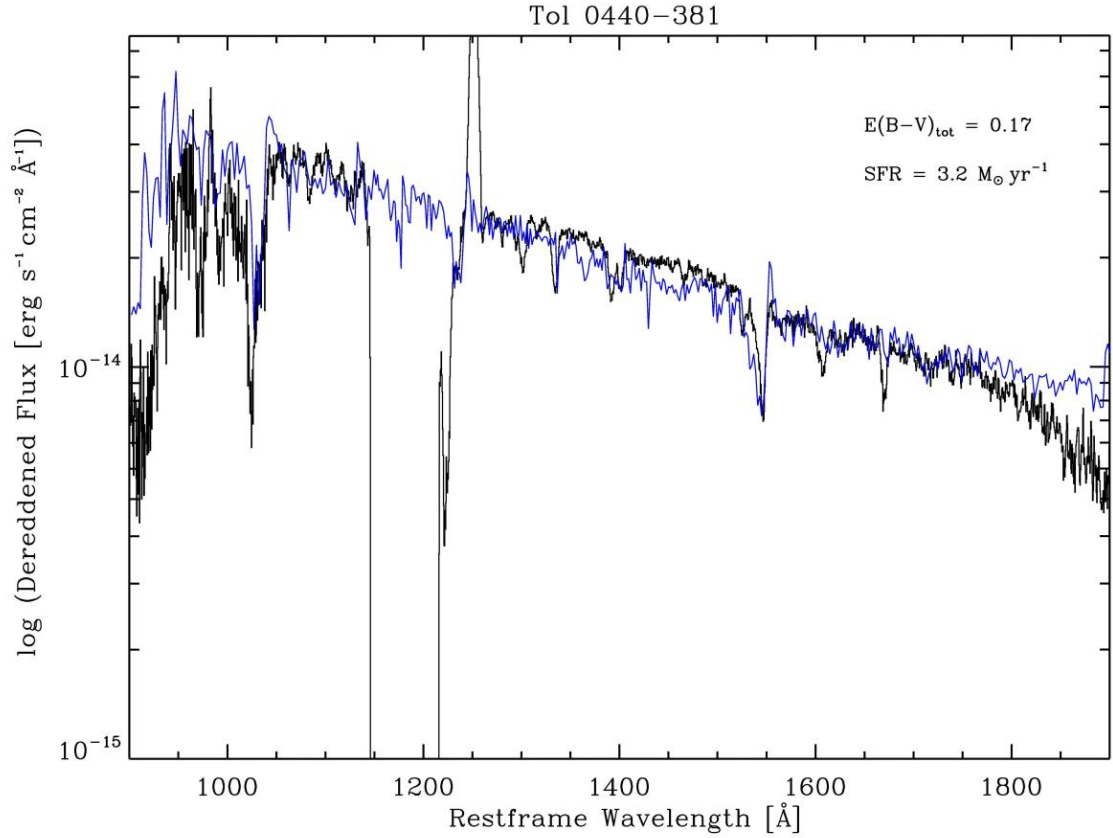

**Figure 4.** Comparison of the observed spectrum of Tol 0440-381 (black) with a synthetic spectrum of a continuously forming population with an age of 20 Myr (blue). See text for additional parameters. The observed spectrum has restframe wavelengths and was corrected for both foreground and intrinsic reddening. The theoretical spectrum was calculated for $D = 167$ Mpc and $SFR = 3.2$ $M_\odot$ yr$^{-1}$.

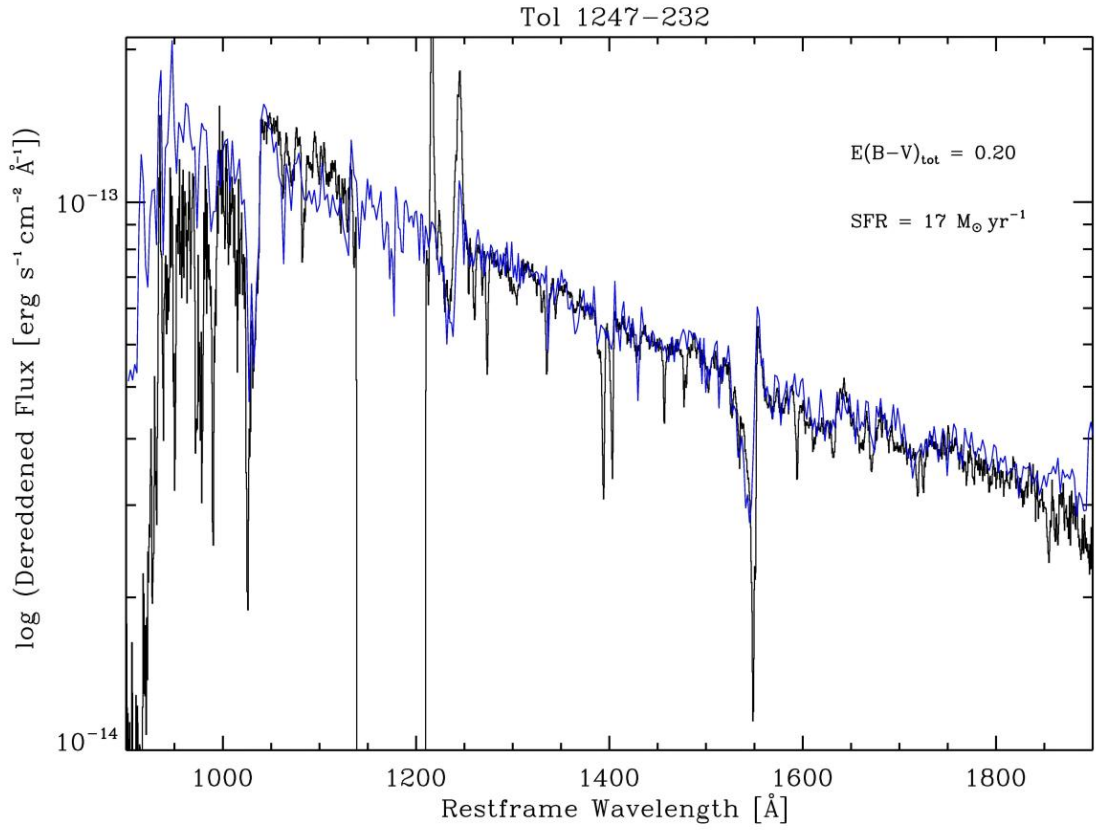

**Figure 5.** Same as Figure 4 but for Tol 1247-232 assuming *D* = 207 Mpc and *SFR* = 17 $M_\odot yr^{-1}$.

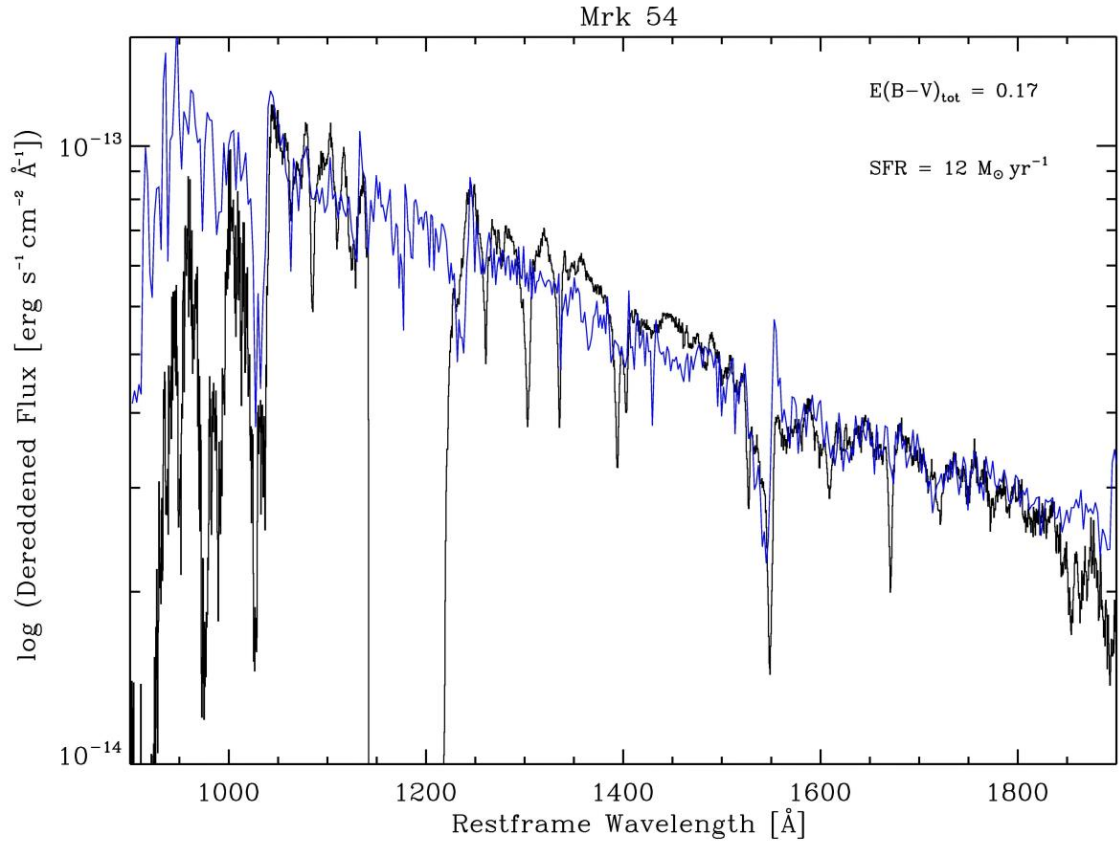

**Figure 6.** Same as Figure 4 but for Mrk 54 assuming $D$ = 191 Mpc and $SFR$ = 12 $M_\odot \mathrm{yr}^{-1}$.

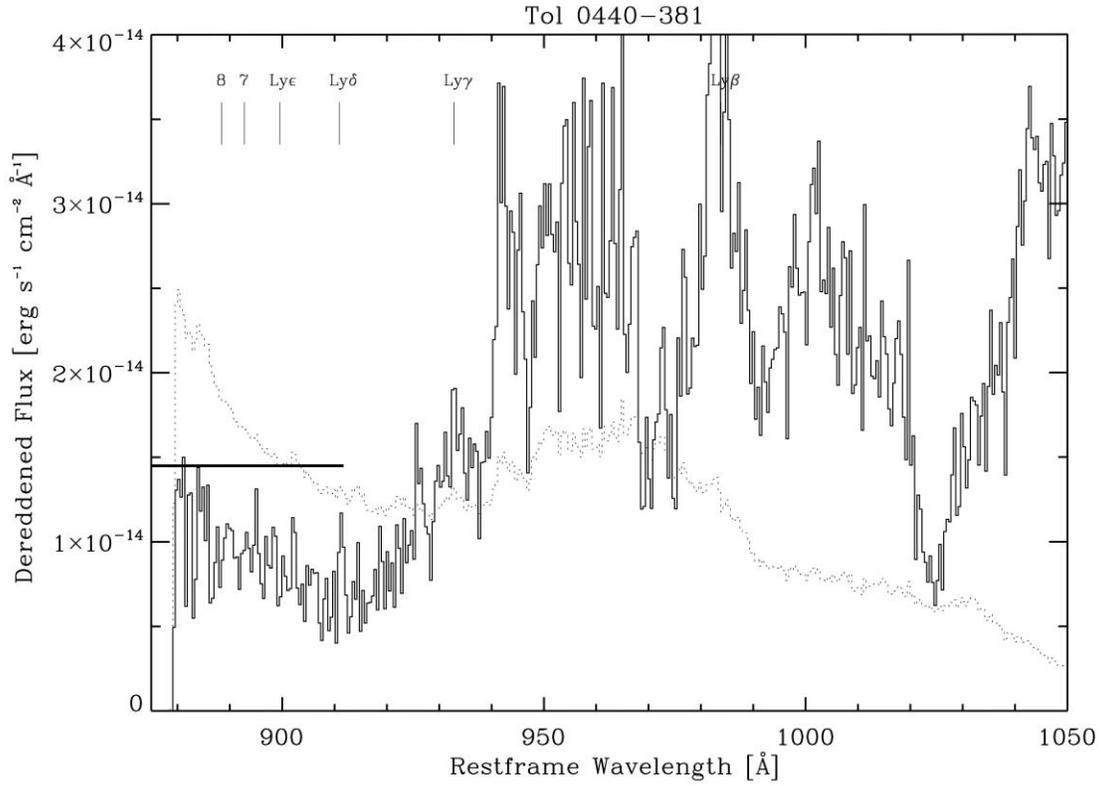

**Figure 7.** Spectral region around the Lyman break for Tol 0440-381. Solid: reddening corrected spectrum; dotted: corresponding 1 σ error vector at the resolution of the binned spectra; horizontal thick bar: location of the predicted Lyman continuum for 100% escape fraction. Tick marks at the top of the figure indicate the wavelengths of geocoronal Lyman lines observed in the restframe of Tol 0440-381.

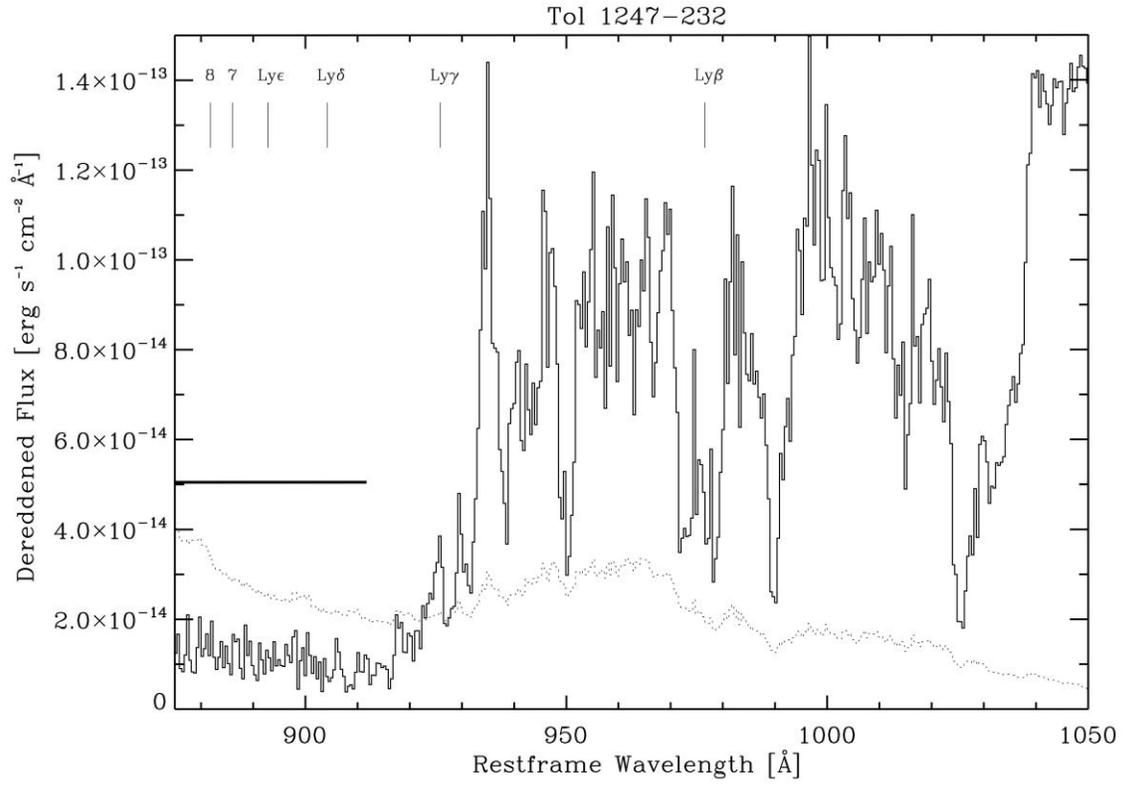

**Figure 8.** Same as Figure 7 but for Tol 1247-232.

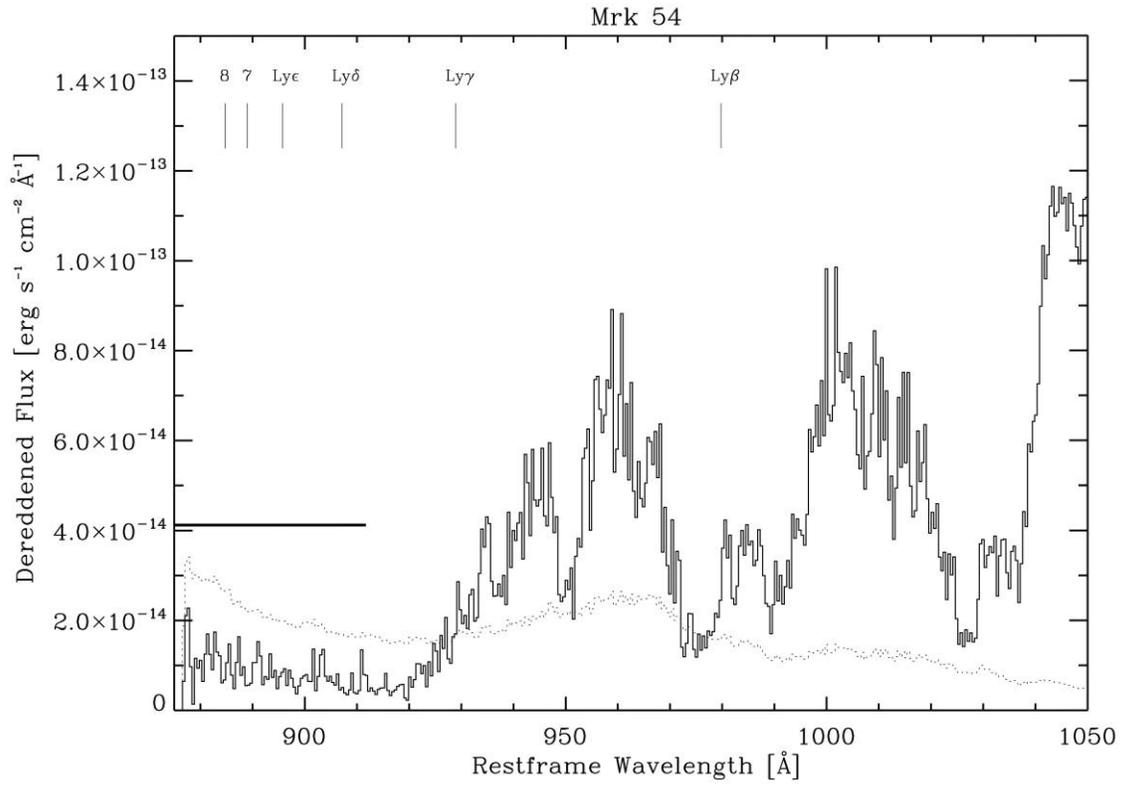

**Figure 9.** Same as Figure 7 but for Mrk 54.

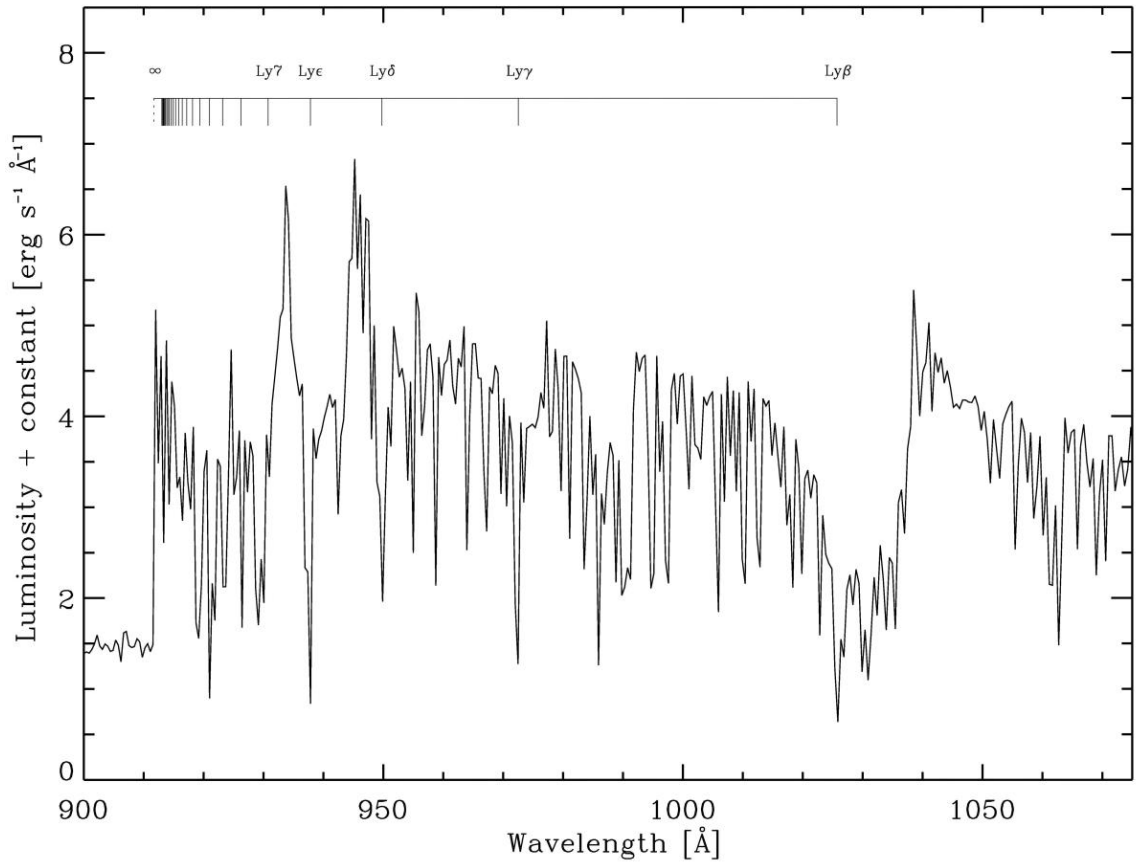

**Figure 10.** Close-up view of the spectral region close to the Lyman break in the synthetic spectrum of the baseline population model. The wavelengths of the Lyman series are marked at the top. WM-Basic does not account for Stark broadening. Therefore the highest Lyman lines are too weak or completely absent. The rise of the spectrum at ∼915 Å results from the missing Lyman blanketing. Prominent lines in the spectrum are O VI $\lambda 1036$ and the S VI doublet at 933 and 945 Å.

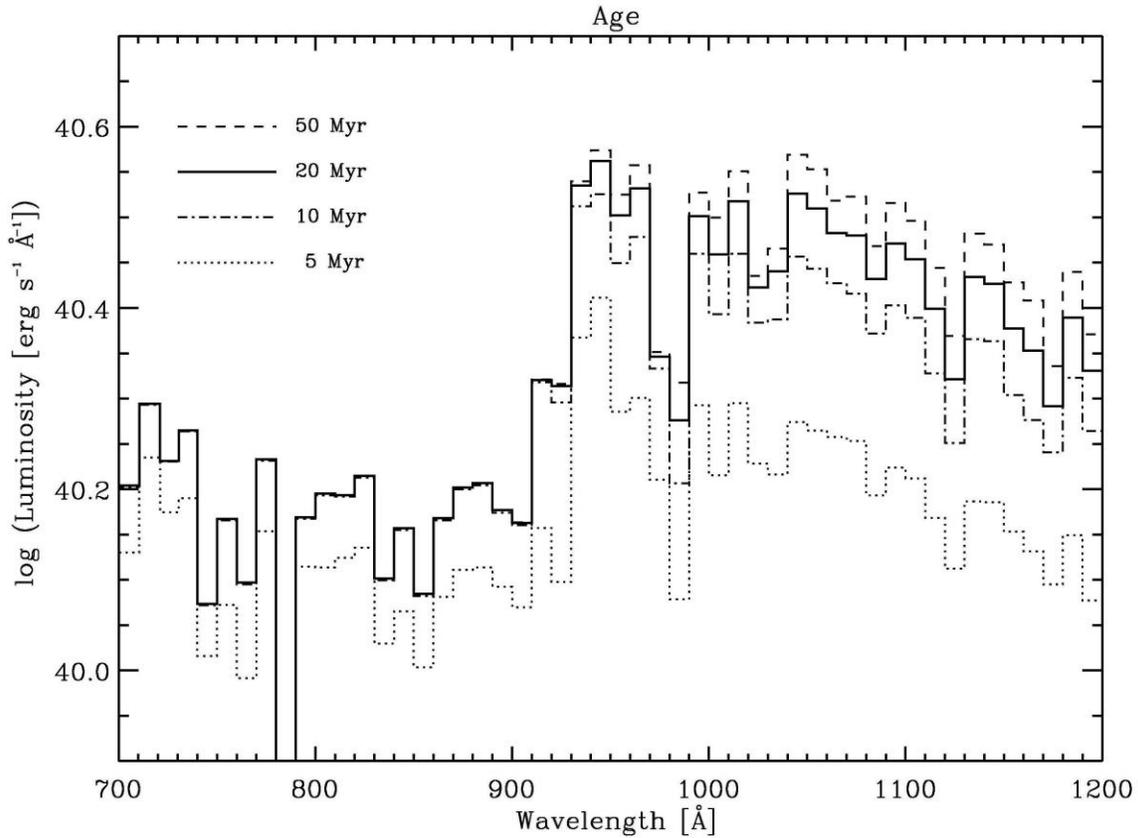

**Figure 11.** Comparison of the predicted Lyman break for models with different ages. Solid: continuous star-formation over a period of 20 Myr. This model was used to derive the Lyman escape fractions in this study. Additional parameters: Kroupa IMF, solar chemical composition, spherically extended, expanding, fully blanketed non-LTE atmospheres, stellar evolution models with rotation. This spectrum is also used as a reference (solid line) in the subsequent figures. Three additional spectra with durations of 5, 10, and 50 Myr are shown. All spectra are binned to a resolution of 10 Å.

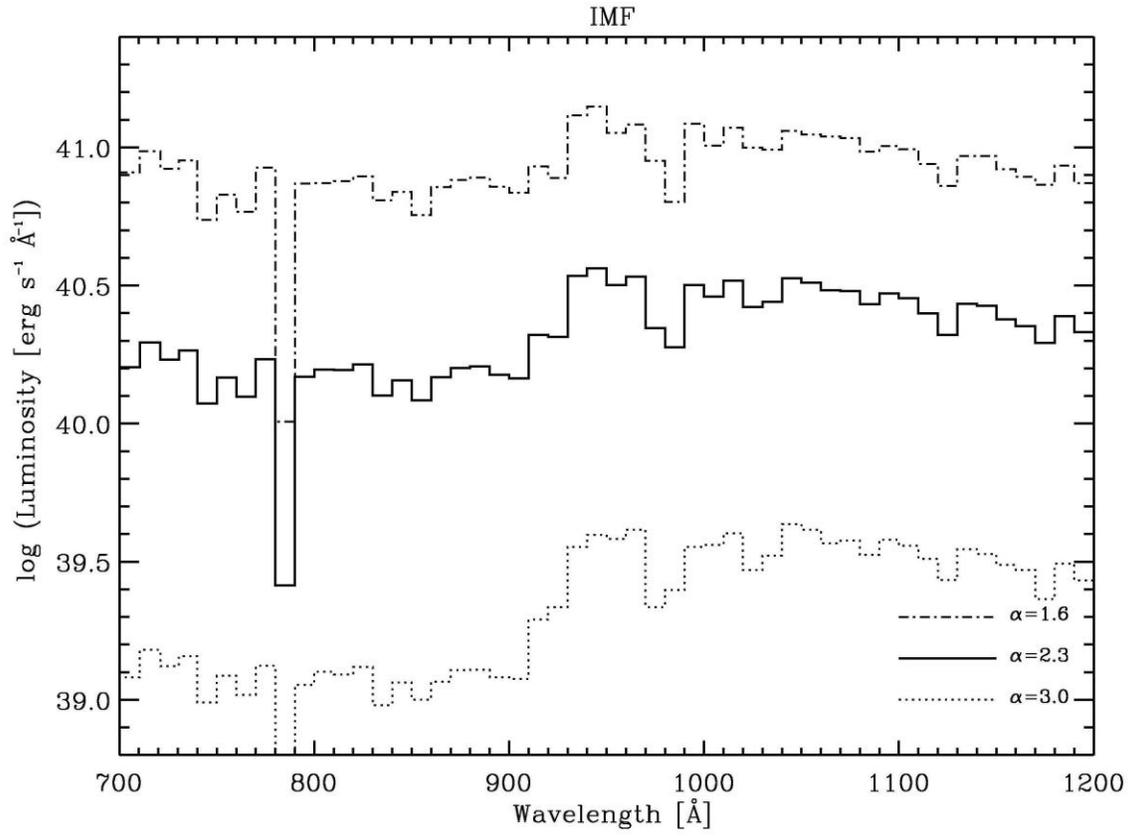

**Figure 12.** Same as Figure 11 but for different IMF exponents at the high-mass end. α = 2.3 corresponds to a Kroupa IMF.

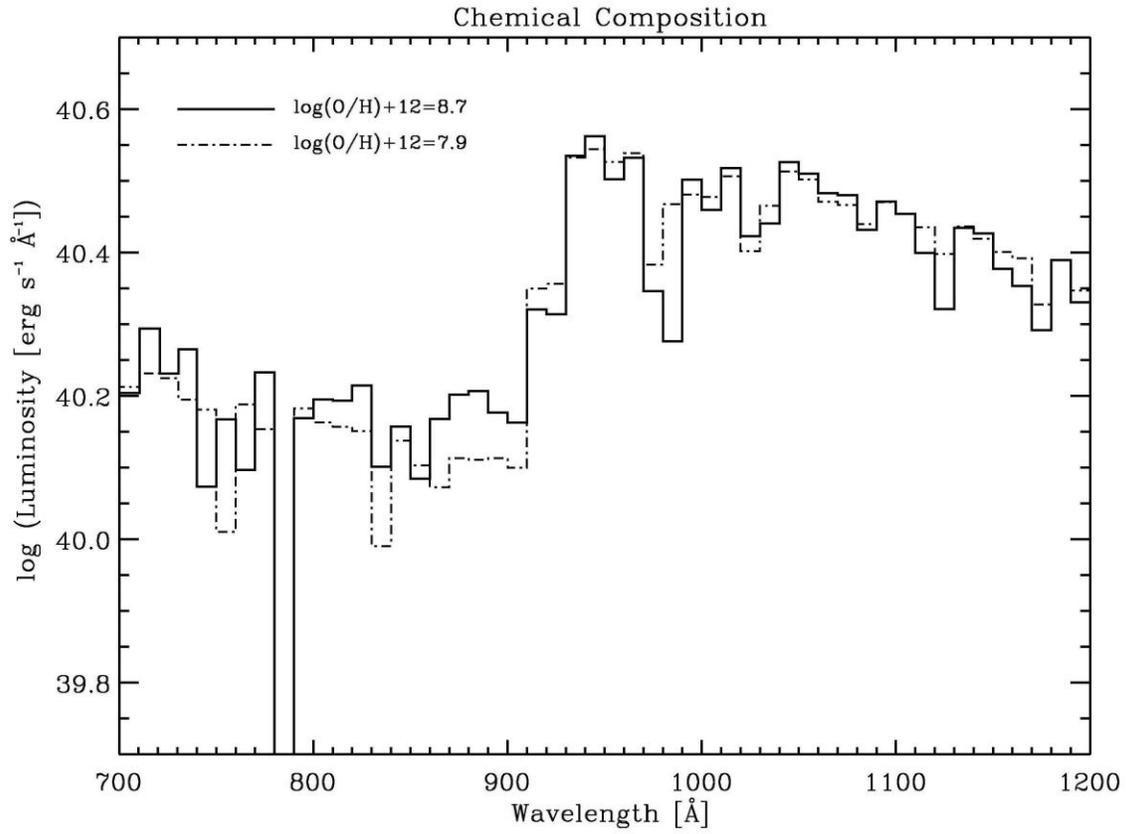

**Figure 13.** Same as Figure 11 but comparing models with heavy-element abundances of solar and 1/7 solar.

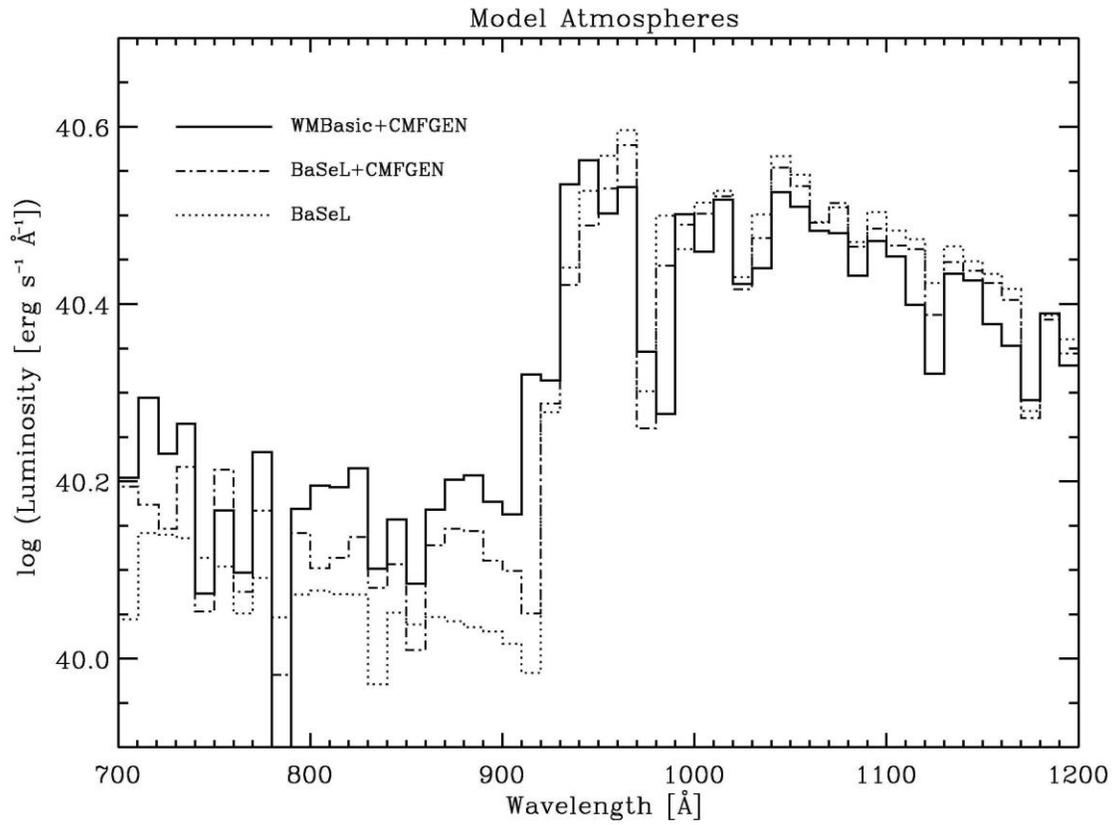

**Figure 14.** Same as Figure 11 but for models with different stellar atmospheres.

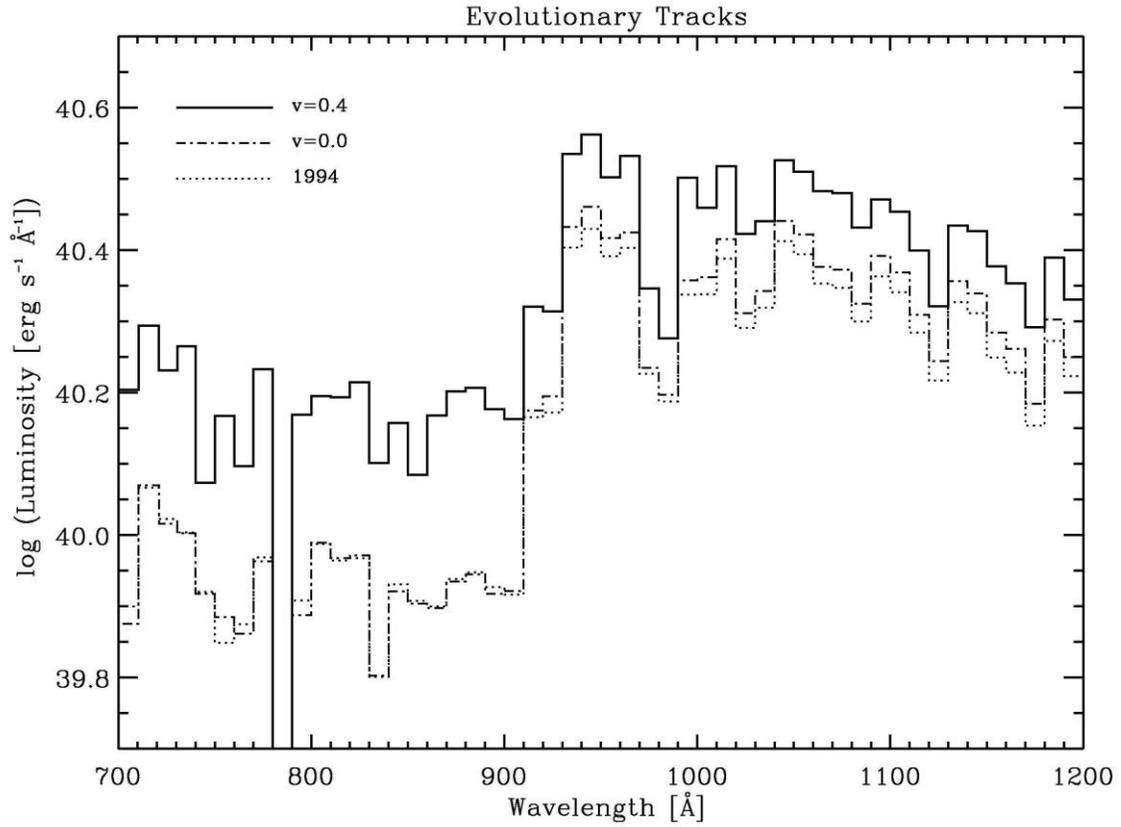

**Figure 15.** Same as Figure 11 but for models with different stellar evolutionary tracks.

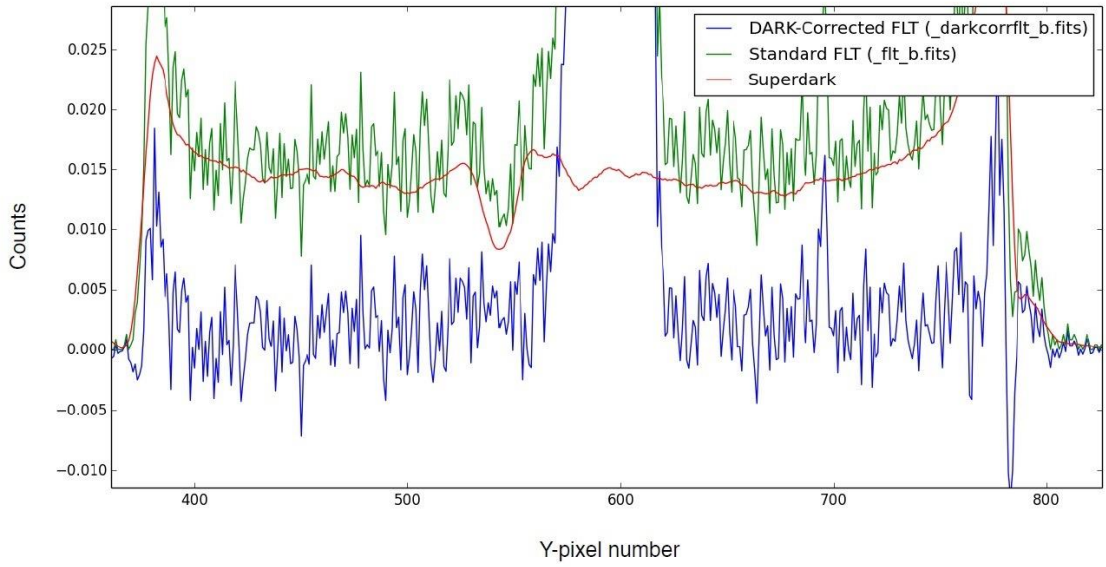

**Appendix Figure 1.** Cross-dispersion profiles of the standard FLT (green), the DARK-corrected FLT in (blue), and the superdark (red). Each profile was obtained by summation over the full x-pixel range.

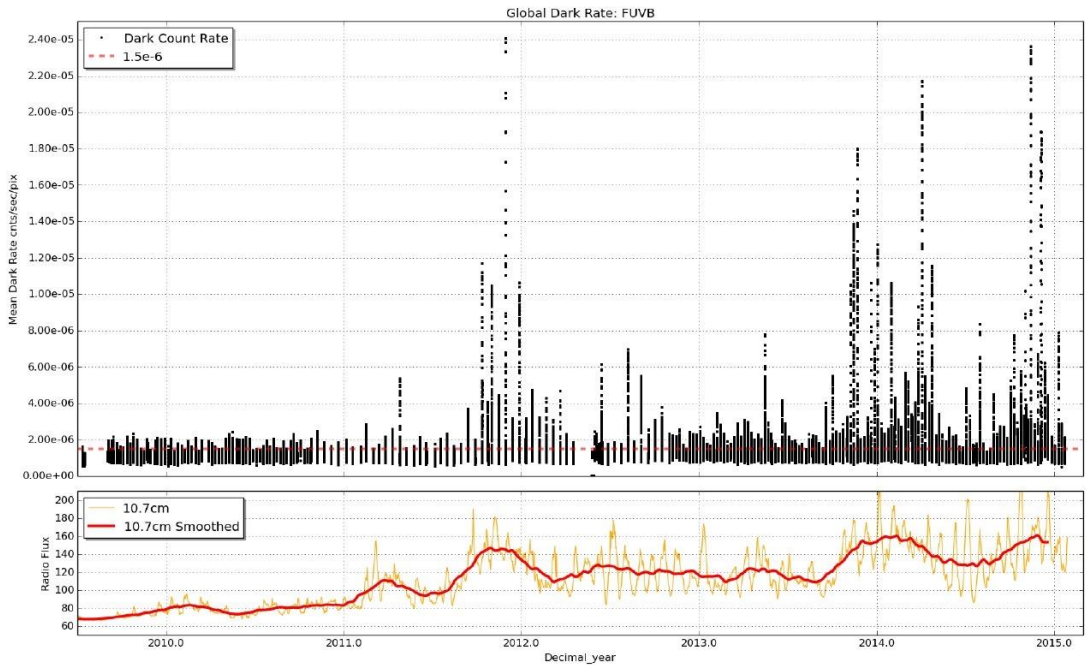

**Appendix Figure 2.** Top panel: COS FUVB global dark rate versus time as measured in the COS FUV Detector Dark Monitor. The dark rate estimates were obtained from all DARK exposures in the archive. Bottom panel: corresponding solar radio emission at 10.7 cm (yellow line), with a smoothed version overplotted (red line).

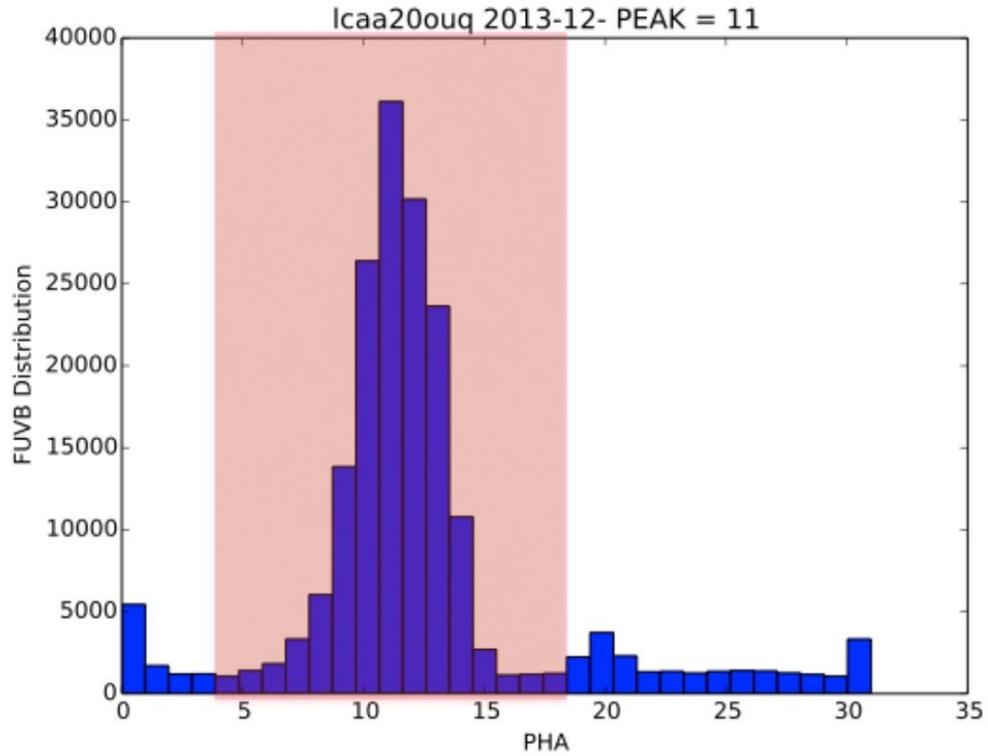

**Appendix Figure 3.** Pulse Height Distribution for COS FUV Segment B science exposure LCAA20OUQ used in selecting new PHA thresholds for calibration. Events outside of the pale-salmon rectangle are considered to be background events, either internal or external.

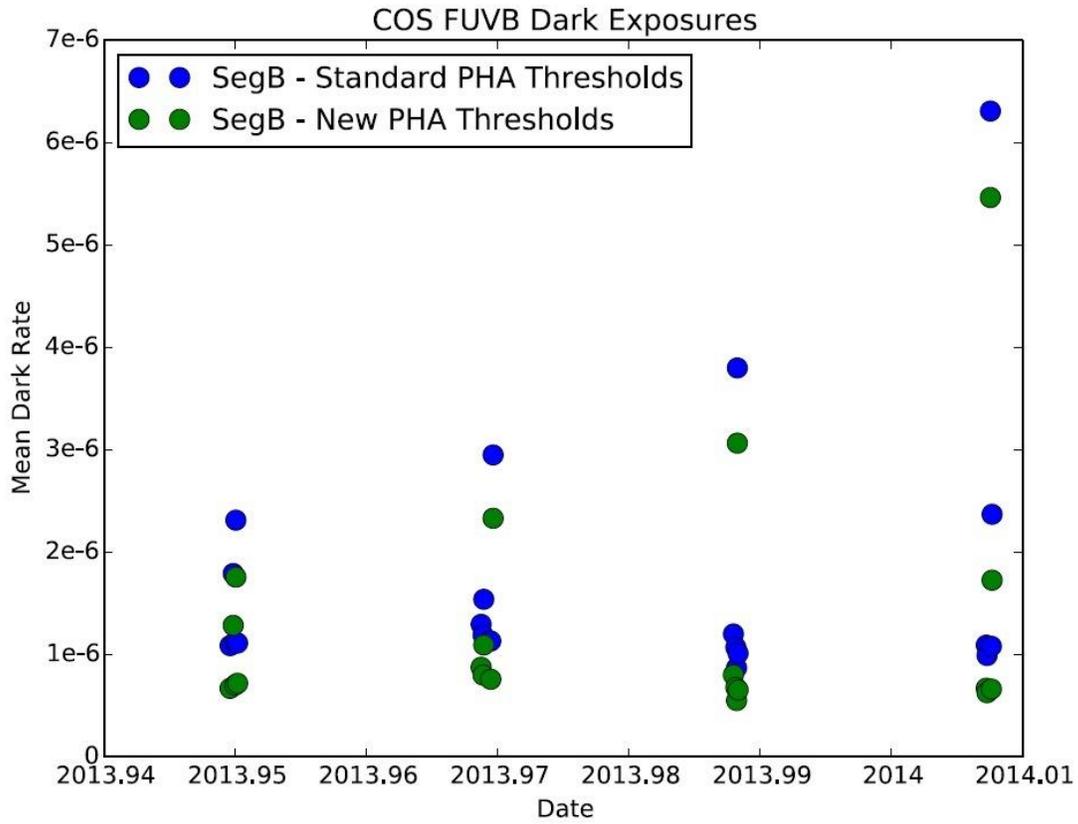

**Appendix Figure 4.** Mean dark rate as a function of time for two sets of dark exposures using different PHA thresholds. Blue: LLT = 2 and ULT = 23; green: LLT = 4 and ULT = 18.

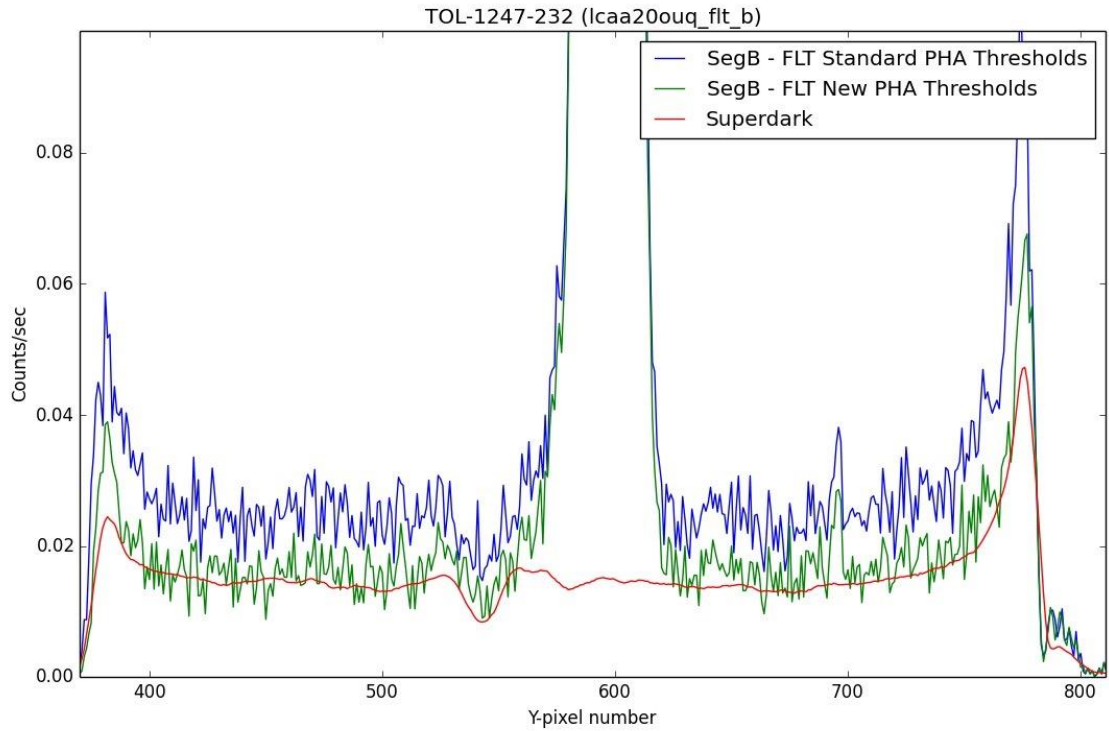

**Appendix Figure 5.** Cross-dispersion profile of the science FLTs. Blue: FLT using the standard PHA thresholds green: FLT with new PHA thresholds; red: the superdark used for the data.

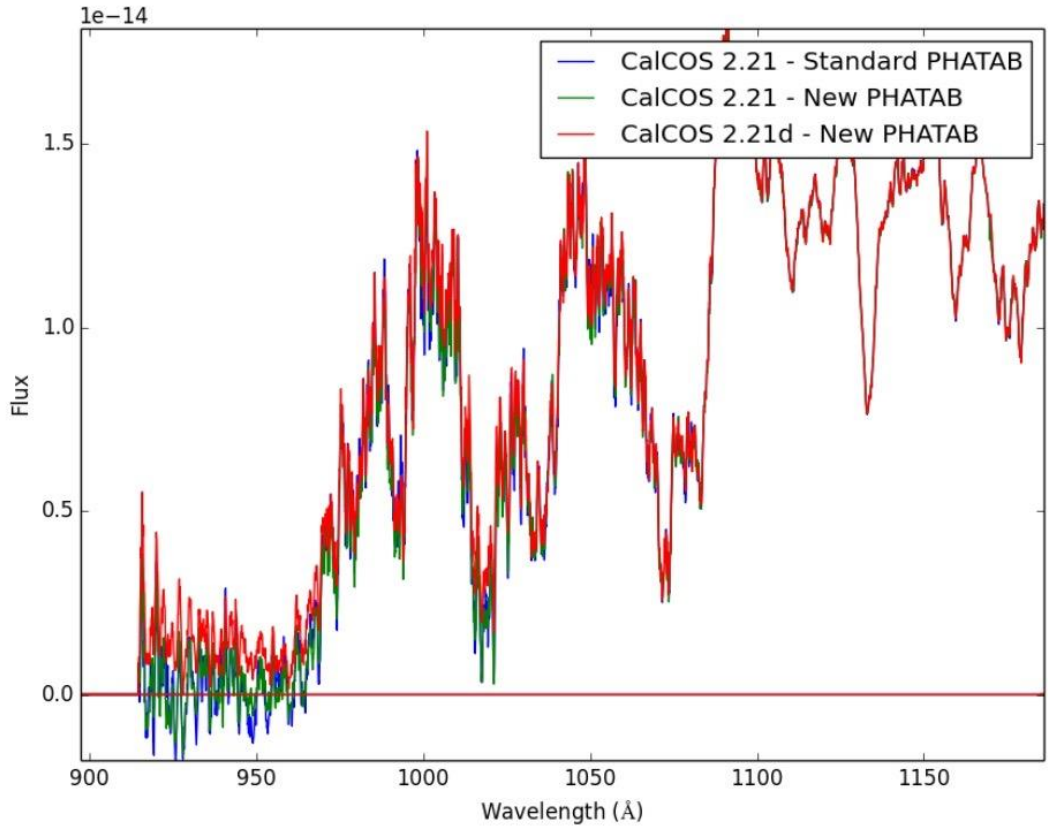

**Appendix Figure 6.** Example of a Segment B calibrated spectrum. Blue: the spectrum calibrated with CalCOS 2.21 and the standard PHATAB (LLT= 2 and ULT=23); green: the spectrum calibrated with CalCOS 2.21 and the new PHATAB (LLT=4 and ULT=18); red: the spectrum calibrated using CalCOS 2.21d and the new PHATAB (LLT=4 and ULT=18).

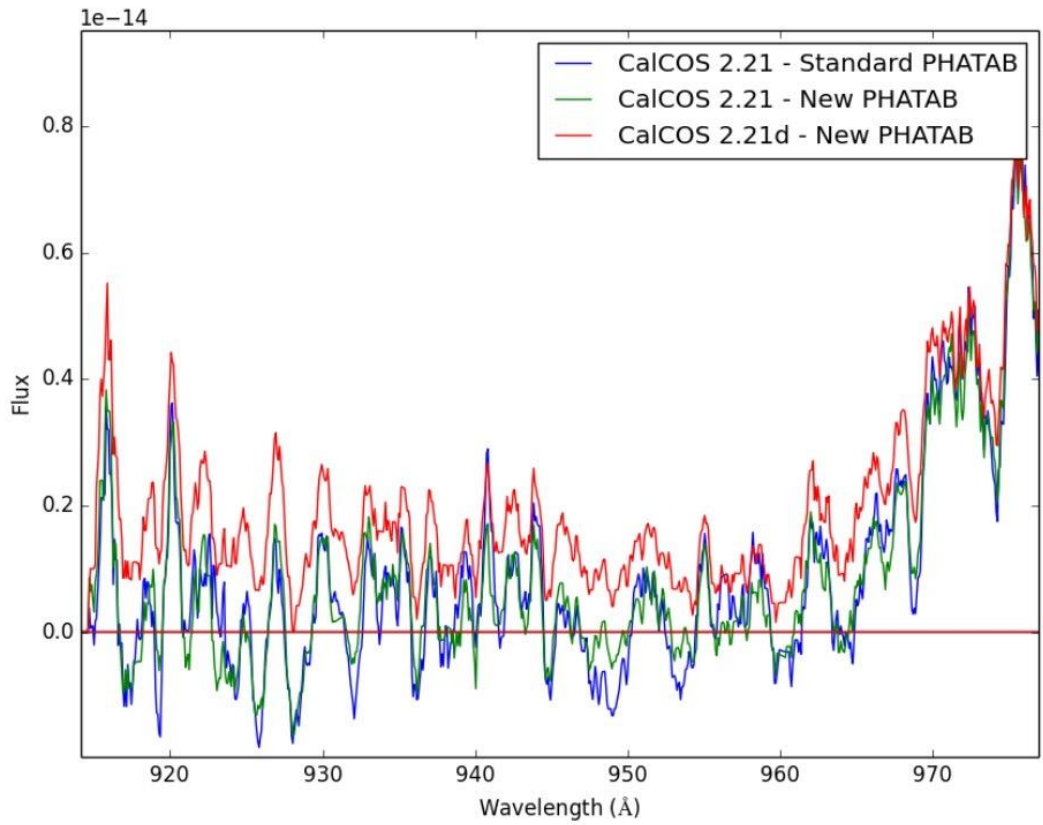

**Appendix Figure 7.** Zoomed version of **Appendix Figure 6**, emphasizing the wavelength region close to the Lyman-break.

# Tables

**Table 1.** Program galaxies

| Galaxy | Classification | $E(B-V)_{MW}$ (mag) | $cz$ (km s$^{-1}$) | $D$ (Mpc) | $M_B$ (mag) | log(O/H)+12 |
|---|---|---|---|---|---|---|
| Tol 0440-381 | HII | 0.016 | 12,270 | 167 | −20.2 | 8.2 |
| Tol 1247-232 | HII | 0.089 | 14,400 | 207 | −21.0 | 8.1 |
| Mrk 54 | BCG | 0.015 | 13,470 | 191 | −22.2 | 8.6 |

**Table 2.** Observation log

| Data Set | Galaxy | RA (J2000) (h m s) | Dec (J2000) (° ′ ″) | Start Time (UT) | Duration (s) |
|---|---|---|---|---|---|
| LCAA10010 | Tol 0440-381 | 04 42 08.10 | −38 01 11.0 | 2014-03-05 05:10 | 11,356 |
| LCAA11010 | Tol 0440-381 | 04 42 08.10 | −38 01 11.0 | 2014-03-06 03:26 | 11,356 |
| LCAA12010 | Tol 0440-381 | 04 42 08.10 | −38 01 11.0 | 2014-03-04 03:40 | 8,004 |
| LCAA51010 | Tol 0440-381 | 04 42 08.10 | −38 01 11.0 | 2014-05-03 02:30 | 14,005 |
| LCAA52010 | Tol 0440-381 | 04 42 08.10 | −38 01 11.0 | 2014-05-12 04:54 | 14,005 |
| LCAA20010 | Tol 1247-232 | 12 50 18.90 | −23 33 57.6 | 2013-12-21 15:34 | 11,184 |
| LCAA21010 | Tol 1247-232 | 12 50 18.90 | −23 33 57.6 | 2013-12-22 17:06 | 11,184 |
| LCAA30010 | Mrk 54 | 12 56 55.66 | +32 26 51.4 | 2014-05-01 11:10 | 11,276 |
| LCAA31010 | Mrk 54 | 12 56 55.66 | +32 26 51.4 | 2014-05-11 13:25 | 7,661 |

**Table 3.** Star-formation properties

| Galaxy | $E(B-V)_{total}$ (mag) | SFR (M$_\odot$ yr$^{-1}$) | $N_{Lyman}$ (s$^{-1}$) | $F_{\lambda,pred}(912^-)$ (erg s$^{-1}$ cm$^{-2}$ Å$^{-1}$) |
|---|---|---|---|---|
| Tol 0440-381 | 0.17 | 3.2 | $7.94 \times 10^{53}$ | $1.45 \times 10^{-14}$ |
| Tol 1247-232 | 0.20 | 17 | $4.21 \times 10^{54}$ | $5.05 \times 10^{-14}$ |
| Mrk 54 | 0.17 | 12 | $2.92 \times 10^{54}$ | $4.12 \times 10^{-14}$ |

**Table 4.** Lyman photon escape fractions

| Galaxy | $F_{\lambda,dered}(912^-)$ (erg s$^{-1}$ cm$^{-2}$ Å$^{-1}$) | $F_{\lambda,obs}(912^-)$ (erg s$^{-1}$ cm$^{-2}$ Å$^{-1}$) | $f_{rel}$ (%) | $f_{abs}$ (%) |
|---|---|---|---|---|
| Tol 0440-381[3] | $(8.68\pm1.99) \times 10^{-15}$ | $(1.03\pm0.23) \times 10^{-15}$ | <59.8±13 | <7.1±1.1 |
| Tol 1247-232 | $(1.09\pm0.30) \times 10^{-14}$ | $(2.26\pm0.62) \times 10^{-15}$ | 21.6±5.9 | 4.5±1.2 |
| Mrk 54 | $(8.25\pm2.51) \times 10^{-15}$ | $(9.76\pm2.94) \times 10^{-16}$ | 20.8±6.1 | 2.5±0.72 |

---

[3] Systematic uncertainties may affect the Lyman continuum fluxes of Tol 0440-381; therefore we consider the escape fractions as upper limits.